\documentclass[prd,preprint,nofootinbib,showpacs]{revtex4}
\usepackage{natbib}
\usepackage{amssymb,amsbsy,amsmath,amsfonts}
\usepackage{graphicx}
\usepackage{float}
\usepackage{rotating}

\def\slashp{p \!\!\! \slash}
\def\slashpm{p' \!\!\!\! \slash}
\def\slashk{k \!\!\! \slash}

\begin{document}
\title {Nucleon-to-Delta axial transition form factors  in relativistic baryon chiral
perturbation theory}

\author{L. S. Geng}
\author{J. Martin Camalich}
\author{L. Alvarez-Ruso}
\author{M. J. Vicente Vacas}

\affiliation{ 
Departamento de F\'{\i}sica Te\'orica and IFIC,
Centro Mixto Universidad de Valencia-CSIC,
Institutos de Investigaci\'on de Paterna, Aptdo. 22085, 46071 Valencia, Spain\\
}

\begin{abstract}
We report a theoretical study of  the axial Nucleon to Delta(1232) ($N\rightarrow\Delta$)
transition form factors up to one-loop order in 
 relativistic baryon chiral perturbation theory.
We adopt a formalism in which the $\Delta$ couplings obey the
spin-3/2 gauge symmetry and, therefore, decouple the unphysical
spin-1/2 fields. 
We compare the results with  phenomenological 
form factors obtained from  neutrino bubble chamber data and in quark models.

\end{abstract}
\date{\today}
 \maketitle

\section{Introduction}
The axial $N\rightarrow\Delta(1232)$ transition form factors play an
important role in neutrino induced pion production on the nucleon,
in particular at low energies~\cite{LlewellynSmith:1971zm,Fogli:1979cz,Sato:2003rq,Hernandez:2007qq,
AlvarezRuso:2007it}. 
These form factors have been
parametrized phenomenologically to fit the
ANL~\cite{Barish:1978pj,Radecky:1981fn} and
BNL~\cite{Kitagaki:1986ct,Kitagaki:1990vs} bubble-chamber data. In the past, the
theoretical descriptions have been done using different
approaches, for a review, see Ref.~\cite{Liu:1995bu}. 
In recent years,
there has been an increasing interest on these form factors. They have
been calculated, for instance, using the
chiral constituent quark model~\cite{BarquillaCano:2007yk} and 
light cone QCD sum rules~\cite{Aliev:2007pi}. State of the art
calculations within lattice QCD
~\cite{Alexandrou:2006mc,Alexandrou:2007zz} have also become
available. The possibility to extract the axial  $N\rightarrow\Delta$ transition form factors 
using parity-violating electron scattering at Jefferson Lab~\cite{Exp:G0}  has been studied extensively~\cite{Mukhopadhyay:1998mn,Zhu:2001br}.
Present and future neutrino experiments could also provide further information on these form factors~\cite{Hasegawa:2005td,
Raaf:2004ty,Wascko:2006ty,Mahn:2006ac,Drakoulakos:2004gn,Ayres:2004js}.

Chiral perturbation theory, based on a simultaneous expansion of QCD
Green functions in powers of the external momenta and of the
quark masses, has achieved remarkable success in describing the dynamics of the
light pseudoscalar mesons at low energies~\cite{Weinberg:1978kz,Gasser:1984gg,Ecker:1994gg,Scherer:2002tk}. The
sector with one baryon is more problematic because, as was shown in
Ref.~\cite{Gasser:1987rb}, the systematic power counting  is lost
since the nucleon mass is not zero in the chiral limit. These
problems were first handled in  heavy baryon chiral
perturbation theory (HB$\chi$PT), where nucleons are treated
semi-relativistically~\cite{Jenkins:1990jv, Bernard:1995dp}. However, in certain cases, this approximation leads to
convergence problems because the Green functions do not satisfy the analytical properties of the fully relativistic theory~\cite{Becher:1999he}.
Recently, the
systematic power counting has also been restored in the relativistic
formulation through either the infrared~\cite{Becher:1999he} or the extended on-mass-shell regularization schemes~\cite{Gegelia:1999qt,Fuchs:2003qc}.

The explicit inclusion of the $\Delta$ in chiral perturbation theory
requires a power counting that properly incorporates the $\Delta$-$N$ mass
difference, $\varDelta\equiv M_\Delta-M_N$, which
is small compared to the chiral symmetry breaking scale.
Two expansion schemes have been proposed. One is the small scale
expansion~\cite{Hemmert:1997ye} which considers $\varDelta$ to be of the same order 
as the other small scales in the theory, i.e., $m_\pi\sim
p\sim\varDelta$. The other is the $\delta$ expansion scheme, which
counts $\varDelta$ differently depending on the energy
domain~\cite{Pascalutsa:2003zk}. Originally, the  small scale expansion
was used in HB$\chi$PT, while recently it has also been implemented in
relativistic chiral perturbation theory~\cite{Bernard:2003xf,Hacker:2005fh}.

The vector $N\rightarrow\Delta$ transition form factors, important
to understand $eN$ ($\gamma N$) reactions and the structure of the
nucleon, have been calculated up to next-to-leading order in both
the small scale expansion HB$\chi$PT~\cite{Gellas:1998wx,Gail:2005gz} and the $\delta$ expansion relativistic baryon $\chi$PT~\cite{Pascalutsa:2005ts,Pascalutsa:2005vq}.
While axial form factors have been addressed in HB$\chi$PT~\cite{Zhu:2002kh}, no calculation has been performed up to now within the relativistic framework. With lattice QCD results becoming
available~\cite{Alexandrou:2006mc}, it is timely to study the axial
transition form factors within relativistic chiral perturbation theory.

In this paper, we use the relativistic baryon chiral perturbation
theory, including explicitly the $\Delta$ resonance, to calculate the
axial $N\rightarrow\Delta$ transition form factors up to order 3 in the $\delta$ expansion. In sect. II, we briefly
explain the power counting, the difference between the small scale expansion
scheme and the $\delta$ expansion scheme, write down the relevant Lagrangians up to next-to-next-to-leading order and the appropriate form of the
$\Delta$ propagator. Loop calculations are
performed in sect. III. In sect. IV, we discuss our results in terms of the
low energy constants and loop functions. In sect. V we
compare the results with both phenomenological parameterizations and other
theoretical calculations. Summary and conclusions are given in sect.
VI.

\section{Power counting, effective Lagrangians, and the $\Delta$ propagator}
\subsection{Power counting}

 A fundamental concept of $\chi$PT (as Effective Field Theory) is the power counting~\cite{Weinberg:1978kz}. It provides a systematic organization of the effective Lagrangians and the corresponding loop-diagrams within a perturbative expansion in powers of $(p/\Lambda_{\chi\mathrm{SB}})^{n_{\chi PT}}$, where $p$ is a small momentum or scale and $\Lambda_{\chi\mathrm{SB}}$, the chiral symmetry breaking scale. In $\chi$PT with pions and nucleons alone the chiral order of a diagram with $L$ loops, $N_\pi$($N_N$) pion (nucleon) propagators, and $V_k$ vertices from $k$th-order Lagrangians is
  \begin{equation}
  n_{\chi PT}=4L-2N_\pi-N_N+\sum_k k V_k\,. \label{eq:counting1}
  \end{equation}
However, in the covariant theory this rule is violated in loops by lower-order analytical pieces~\cite{Gasser:1987rb}. This power counting can be recovered by adopting non-trivial renormalization schemes, where the lower-order power-counting breaking pieces of the
loop results are systematically absorbed into the available counter-terms~\cite{Becher:1999he,Fuchs:2003qc}.
A detailed discussion of the renormalization scheme adopted in the present work
will be presented together with our main results in section IV.

If the $\Delta$ resonance is explicitly considered, things become more complicated because its 
excitation energy, $\varDelta \equiv M_\Delta-M_N\sim0.3$\,GeV, is
 small compared to the chiral symmetry breaking scale
$\Lambda_{\chi\mathrm{SB}}=4\pi f_\pi\sim 1$\,GeV. Therefore, there are
 two  small parameters in the theory, i.e.,
  \begin{equation}
  \varepsilon=m_\pi/\Lambda_{\chi\mathrm{SB}}\quad\mbox{and}\quad
  \delta=\varDelta/\Lambda_{\chi\mathrm{SB}}.
  \end{equation}

Over the past few years, two different expansion schemes have been
proposed, the small scale expansion and the $\delta$ expansion.
In the small scale expansion~\cite{Hemmert:1997ye}, one has $m_\pi\sim\varDelta\sim
p\sim\mathcal{O}(\epsilon)$. In the $\delta$-expansion~\cite{Pascalutsa:2003zk}, to maintain
the scale hierarchy $m_\pi\ll\varDelta\ll\Lambda_{\chi\mathrm{SB}}$,
 $m_\pi/\Lambda_{\chi\mathrm{SB}}$ is counted as $\delta^2$. In this scheme, the power
counting depends on the energy domain under study:
$p\sim m_\pi$ or $p\sim\varDelta$.

For the study of $N\rightarrow\Delta$ axial transition form factors in the energy region $p\sim\varDelta$,
the order of  a graph with $L$ loops, $V_k$ vertices of dimension $k$,
$N_\pi$ pion propagators, $N_N$ nucleon propagators, $N_\Delta$
Delta propagators, the power-counting index
$n$ is given by:
  \begin{equation}
  n=n_{\chi\mathrm{PT}}-N_\Delta.
 \end{equation}
For a more general discussion, see Ref.~\cite{Pascalutsa:2006up}.

In the present work, we adopt the $\delta$ expansion scheme. As can be seen in the following sections,
the differences between these two schemes in our case come from vertices proportional 
to $m_\pi^2$, which count as $\delta^4$ in the $\delta$ expansion and, therefore, have been
neglected.
\subsection{Chiral Lagrangians}
In this section, we write down the relevant $N N$, $N\Delta$, and $\Delta\Delta$
Lagrangians and pay special attention to the $\Delta$ couplings and the spin-3/2 gauge 
symmetry.
\subsubsection{Pion-nucleon and pion-pion Lagrangians}
The lowest order pion-nucleon Lagrangian has the following form:
 \begin{equation}
 \mathcal{L}^{(1)}_{\pi N}=\bar{N}(i\gamma^\mu
 D_\mu-M_N-\frac{g_A}{2}\gamma^\mu\gamma^5 u_\mu)N,
 \end{equation}
where $M_N$ and $g_A$ are the nucleon mass and the axial-vector coupling at the chiral limit,
$D_\mu$ is the covariant derivative 
 \begin{equation}
 D_\mu N =\partial_\mu N +[\Gamma_\mu, N],
 \end{equation}
 \begin{equation}
 \Gamma_\mu=\frac{1}{2}\left\{u^\dagger(\partial_\mu-i r_\mu)
u+u(\partial_\mu-il_\mu)u^\dagger\right\},
 \end{equation}
and $u_\mu$ the axial current defined as
 \begin{equation}
 u_\mu=i\left\{u^\dagger(\partial_\mu-ir_\mu)u-u(\partial_\mu-i
 l_\mu) u^\dagger\right\}.
 \end{equation}
In the above definitions, $r_\mu=v_\mu+a_\mu$, $l_\mu=v_\mu-a_\mu$ with
$v_\mu=\tau^\sigma v^\sigma_\mu/2$ and $a_\mu=\tau^\sigma a^\sigma_\mu/2$ the external vector and axial currents,
where $\tau^\sigma$ are the Pauli matrices.
The matrix $u$ incorporates the pion fields 
 \begin{equation}
u^2=U=\exp\left[i\frac{\Phi}{f_\pi}\right],
\end{equation}
\begin{equation}
\Phi=\tau_\sigma\pi^\sigma=\left(\begin{array}{cc}\pi^0&\sqrt{2}\pi^+\\
                             \sqrt{2}\pi^-&-\pi^0\end{array}\right),
\end{equation}
with $f_\pi$ being the pion decay constant in the chiral limit.

The leading order 
pion-pion Lagrangian has the following form:
\begin{equation}
\mathcal{L}^{(2)}_{\pi\pi}=\frac{f_\pi^2}{4}\mathrm{Tr}\left[\nabla_\mu U (\nabla^\mu
U)^\dagger\right]+\frac{f_\pi^2}{4}\mathrm{Tr}\left[\chi U^\dagger +
U\chi^\dagger\right],
\end{equation}
where
\begin{equation}
\nabla_\mu U=\partial_\mu U -i r_\mu U +i U l_\mu
\end{equation}
with $\chi=\mathrm{diag}(m^2_\pi,m^2_\pi)$. 
\subsubsection{Nucleon-Delta and Delta-Delta Lagrangians}
 The $\Delta(1232)$ is a spin-3/2 resonance and, therefore, its spin
content can be described in terms of the Rarita-Schwinger (RS) field 
$\Delta_\mu$, where $\mu$ is the Lorentz index.\footnote{
We follow Ref.~\cite{Pascalutsa:2006up} and
write the Lagrangians for the spin-3/2 isospin-3/2 $\Delta$ isobar in
terms of the Rarita-Schwinger (vector-spinor) isoquartet field
$\Delta_\mu=(\Delta^{++},\Delta^{+},\Delta^0,\Delta^{-})_\mu^t$,
which is connected to the isospurion representation of
Ref.~\cite{Hemmert:1997ye} through
\begin{equation*}
\Delta^a_\mu=-T^a\Delta_\mu
\end{equation*}
where $T^a$ are the isospin 1/2 to 3/2 matrices satisfying $T^a T^{b\dagger}=\delta^{ab}-1/3\tau^a\tau^b$, as given in Appendix A.
With this rule, the on-shell
equivalent form of our {\it consistent} couplings can be easily
identified with those of Refs.~\cite{Hemmert:1997ye,Fettes:2000bb}.}
This field, however, contains unphysical spin-1/2 components. They are allowed for
	the description of off-shell Delta's, but the physical results should not depend on them. In order to tackle this problem,
 we follow Refs.~\cite{Pascalutsa:2000kd,Pascalutsa:2006up} and adopt the
{\it consistent} couplings, which are gauge-invariant under the
transformation
\begin{equation}
\Delta_\mu(x)\rightarrow\Delta_\mu(x)+\partial_\mu\epsilon(x).
\end{equation}
 A remarkable  consequence of the use of the spin-3/2
gauge symmetric couplings is that it leads to a natural decoupling
of the propagation of the spin-1/2 fields.

In the following we give the $N\Delta$ and $\Delta\Delta$ 
Lagrangians relevant to this work. 
The lowest order Lagrangians in the resonance region are\footnote{If
one $\Delta$ is put on-shell, the $\Delta$-$\Delta$ Lagrangian is equivalent to that of Pascalutsa et al.~\cite{Pascalutsa:2006up}:
\begin{equation*}
\mathcal{L}_{\Delta\Delta}^{(1)}=-\frac{H_A}{2M_\Delta}\epsilon^{\mu\nu\rho\sigma}\bar{\Delta}_\mu
\mathcal{T}^a(\partial_\rho\Delta_\nu)\omega_\sigma^a+\mathrm{H.c.}.
\end{equation*}
 }
\begin{equation}
\mathcal{L}_{N\Delta}^{(1)}=-\frac{ih_A}{2
M_\Delta}\bar{N}T^a\gamma^{\mu\nu\lambda}(\partial_\mu\Delta_\nu)\omega^a_\lambda
+\mathrm{H.c.},
\end{equation}
\begin{equation}
\mathcal{L}_{\Delta\Delta}^{(1)}=\frac{H_A}{2 M^2_\Delta}\partial_m\bar{\Delta}_b\gamma^{\ell b m}\gamma^\mu\gamma^5
\mathcal{T}^a\omega^a_\mu\gamma_{\ell c n}\partial^n\Delta^c,
\end{equation}
where
$\omega^a_\lambda=\frac{1}{2}\mathrm{Tr}\left(\tau^a u_\lambda\right)=-\frac{1}{f_\pi}\partial_\lambda\pi^a+a^a_\lambda+\cdots$,
$T^a$ and $\mathcal{T}^a$ are the isospin 1/2 to 3/2 and 3/2 to
3/2 transition matrices, and $\gamma^{\mu\nu\lambda}$ is the totally antisymmetric gamma matrix product as given in Appendix A.
 At second
order, there are four terms, i.e.,\footnote{In our study of the axial form factors up to
one-loop order the $\delta^{(2)}$ and $\delta^{(3)}$ Lagrangians only concern on-shell $\Delta$'s. Therefore, they are the same in the {\it consistent} coupling scheme of Pascalutsa et al. as those conventional Lagrangians in Refs.~\cite{Hemmert:1997ye,Fettes:2000bb}.}
\begin{eqnarray}
\mathcal{L}_{N\Delta}^{(2)}&=&-  \frac{d_1}{M_\Delta}\bar{N}T^a
(\partial_\mu\Delta_\nu) f_-^{a,\mu\nu}
-id_2\bar{N}T^af_-^{a,\mu\nu}\gamma_\mu\Delta_\nu
-id_3\bar{N}T^a\omega^{a,\mu\nu}\gamma_\mu\Delta_\nu\nonumber\\
&&
- \frac{d_4}{M_\Delta}\bar{N}T^a
 (\partial_\mu\Delta_\nu)\omega^{a,\mu\nu}+\mathrm{H.c.},
\end{eqnarray}
 while at third order, there are seven terms\footnote{
In the small scale expansion scheme, there are two more terms at this order proportional to $m_\pi^2$, i.e.,
\begin{equation*}
 -f_8 \bar{N}T^a \omega_\nu^a\mathrm{Tr}[\chi_+]\Delta^\nu-f_9 i
\bar{N}T^a \left[D_\nu,\chi^a_-\right]\Delta^\nu,
\end{equation*}
where $\chi_+$ and $\chi_-$ are external scalar and pseudoscalar sources.}
\begin{eqnarray}
\mathcal{L}_{N\Delta}^{(3)}&=&- f_1\bar{N}T^a
\Delta_\nu\partial_\mu f_-^{a,\mu\nu} - f_2
\bar{N}T^a \Delta_\nu\partial_\mu \omega^{a,\mu\nu}+i\frac{f_3}{M_\Delta} \bar{N}T^a\partial^\mu f_-^{a,\alpha\nu}\gamma_\nu\partial_\mu\Delta_\alpha\nonumber\\
&&+
i\frac{f_4}{M_\Delta}\bar{N}T^a\partial^\mu f_-^{a,\alpha\nu}\gamma_\mu\partial_\nu\Delta_\alpha
-i\frac{f_5}{M_\Delta}\bar{N}T^a\partial^\mu\omega^{a,\nu\alpha}\gamma_\mu\partial_\nu\Delta_\alpha\nonumber\\
&&
+\frac{f_6}{M^2_\Delta}\bar{N}T^a\partial^\mu f^{a,\nu\alpha}_-\partial_\mu\partial_\nu\Delta_\alpha
+\frac{f_7}{M^2_\Delta}\bar{N}T^a\partial^\mu\omega^{a,\nu\alpha}\partial_\mu\partial_\nu\Delta_\alpha
 + \mathrm{H.c.},
\end{eqnarray}
where
$\omega^a_{\mu\nu}=\frac{1}{2}\mathrm{Tr}(\tau^a[D_\mu,u_\nu])$,
$f_-^{a,\mu\nu}=\partial^\mu a^{a,\nu}-\partial^\nu a^{a,\mu}$.
As we will see later, the $\delta^{(2)}$ and $\delta^{(3)}$ low-energy constants (LEC) contribute to the form factors only in particular combinations; therefore, the number of independent parameters is smaller than the one 
appearing in the above Lagrangians.

\subsection{Spin-3/2 propagator}
 The most general spin-3/2 free
field propagator in $D$ dimensions has the following form~\cite{Bernard:2003xf,Pascalutsa:2005nd}:
 \begin{eqnarray}\label{eq:3halfprop}
 S^{\alpha\beta}(p)&=&\frac{\slashp+M_\Delta}{M_\Delta^2-p^2}\bigg[
 g^{\alpha\beta}-\frac{\gamma^\alpha\gamma^\beta}{(D-1)}
+\frac{(1-\zeta)(\zeta\slashp+M_\Delta)}{(D-1)(\zeta^2
 p^2-M_\Delta^2)}(\gamma^\alpha p^\beta-\gamma^\beta
 p^\alpha)\nonumber\\
&&\hspace{8cm}+\frac{(D-2)(1-\zeta^2)p^\alpha p^\beta}{(D-1)(\zeta^2
 p^2-M_\Delta^2)}\bigg],
 \end{eqnarray}
where $\zeta$ is the spin-3/2 gauge-fixing parameter.
 In the case of $\zeta=0$, the above propagator corresponds to
 the usual Rarita-Schwinger propagator
 \begin{equation}
 S^{\alpha\beta}(p)=\frac{\slashp+M_\Delta}{M_\Delta^2-p^2}
 \left[g^{\alpha\beta}-\frac{\gamma^\alpha\gamma^\beta}{(D-1)}-\frac{1}{(D-1)M_\Delta}(\gamma^\alpha p^\beta-\gamma^\beta
 p^\alpha)-\frac{(D-2)p^\alpha p^\beta}{(D-1)M_\Delta^2} \right];
 \end{equation}
  while in the case of
 $\zeta=\infty$, it becomes
 \begin{equation}
 S^{\alpha\beta}(p)=\frac{\slashp+M_\Delta}{M_\Delta^2-p^2}\mathcal{P}^{\alpha\beta}_{3/2}(p)
 \end{equation}
 with the covariant spin-3/2 projection operator defined by
 \begin{equation}
\mathcal{P}^{\alpha\beta}_{3/2}(p)=g^{\alpha\beta}-\frac{\gamma^\alpha\gamma^\beta}{(D-1)}-\frac{1}{(D-1)p^2}
\left(\slashp\gamma^\alpha
p^\beta+p^\alpha\gamma^\beta\slashp\right)-\frac{(D-4)p^\alpha p^\beta}{(D-1)p^2}.
 \end{equation}

It should be stressed that due to the spin-3/2 gauge symmetric nature of
the {\it consistent} couplings, our results do not depend on the
particular value of the gauge-fixing parameter $\zeta$.
\begin{figure}[t]
\includegraphics[scale=0.7]{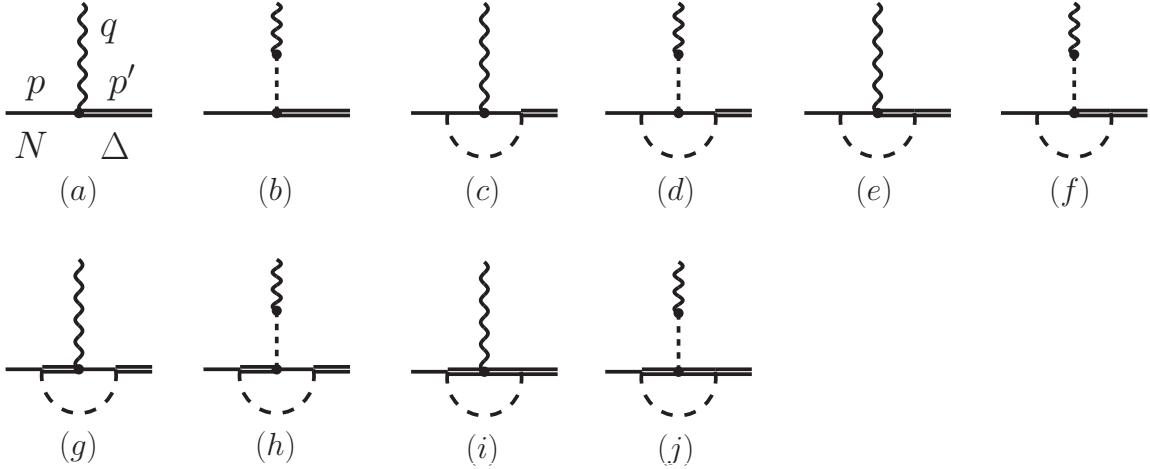}
\caption{Feynman diagrams contributing to the axial $N\rightarrow\Delta$
transition form factors up to $\delta^{(3)}$. The double, solid, and
dashed lines correspond to the Delta, nucleon, and pion, while the
wiggly line denotes the external axial
source. \label{fig_diagram}}
\end{figure}
\section{The $N\rightarrow\Delta$ axial transition form factors}
The $N\rightarrow\Delta$ axial transition form factors can be parameterized in
terms of the usually called Adler form factors~\cite{LlewellynSmith:1971zm,Schreiner:1973mj}:
\begin{eqnarray}\label{Adlerff}
\langle\Delta^+_{\alpha}(p')|-A^{\alpha\mu,3}|P(p)\rangle&=&
 \bar{\Delta}^+_{\alpha}(p')\bigg\{
 \frac{C^A_3(q^2)}{M_N}\big(g^{\alpha\mu}\gamma\cdot
 q-q^\alpha\gamma^\mu\big)+
 \frac{C^A_4(q^2)}{M^2_N}\big(q\cdot p' g^{\alpha\mu}- q^\alpha p'^\mu
\big)\nonumber\\
 &&\hspace{3cm}+C^A_5(q^2)g^{\alpha\mu}+\frac{C^A_6(q^2)}{M^2_N}q^\alpha
 q^\mu\bigg\}N,
\end{eqnarray}
where $A^{\alpha\mu,3}$ is the third isospin component of the axial current.

 All the diagrams contributing to the $N\rightarrow\Delta$ axial transition
 form factors up to $\delta^{(3)}$  are displayed in
 Fig.~\ref{fig_diagram}.\footnote{
 We do not have the diagrams (c), (d), (e), and (f) of Fig.~1 of Ref.~\cite{Zhu:2002kh}, that
 correspond to tadpole diagrams where a pion loop couples to either
 the $A N\Delta$ ($\pi N\Delta$, $A\pi$) vertices, or the pion fields, because
the contribution of those diagrams are of higher-order in the $\delta$ expansion scheme.
} Two Kroll-Ruderman like diagrams are not shown since the one with an internal nucleon and  a $A\pi NN$ vertex is zero and
the other one with an internal $\Delta$ and a $A\pi\Delta\Delta$ vertex contributes as a real constant, which is irrelevant to
the present study due to the adopted renormalization scheme.
The calculation of the tree-level diagrams [Fig.~1(a)] is straightforward:
\begin{eqnarray}
 A^{\alpha\mu,3}_{(a)}&=&\sqrt{\frac{2}{3}}\Bigg[-\frac{h_A}{2} g^{\alpha\mu}
- \frac{d_1}{M_\Delta} (p'\cdot
qg^{\alpha\mu}-q^\alpha p'^\mu)  
-d_2
(\gamma\cdot q g^{\alpha\mu}-q^\alpha\gamma^\mu)-d_3\gamma\cdot q g^{\alpha\mu}\nonumber\\
&&\hspace{1.3cm}-\frac{d_4}{M_\Delta}p'\cdot q g^{\alpha\mu}
+f_1 ( q^2
g^{\alpha\mu}-q^\alpha q^\mu)+ f_2 q^2 g^{\alpha\mu}
+\frac{f_3}{M_\Delta} p'\cdot q(\gamma\cdot q g^{\alpha\mu}-q^\alpha\gamma^\mu)\nonumber\\
&&\hspace{1.3cm}+ \frac{f_4}{M_\Delta} \gamma\cdot q\left(p'\cdot q g^{\alpha\mu}-q^\alpha p'^\mu \right)
+\frac{f_5}{M_\Delta}\gamma\cdot q p'\cdot q g^{\alpha\mu}
+\frac{f_6}{M^2_\Delta}p'\cdot q(p'\cdot qg^{\alpha\mu}-q^\alpha p'^\mu)\nonumber\\
&&\hspace{1.3cm}+\frac{f_7}{M^2_\Delta} p'\cdot q p'\cdot q g^{\alpha\mu}\Bigg],
\end{eqnarray}
where $p'$, $p$, and $q$ are the momenta of the $\Delta$,  the nucleon, and 
the external source. We assume that both the external nucleon and $\Delta$ are on-shell, which yields 
$p'\cdot q/M_\Delta= (M^2_\Delta-M^2_N+q^2)/(2M_\Delta)\approx\varDelta$ and $\gamma\cdot q=M_\Delta-M_N=\varDelta$, where we have neglected the $q^2$ and $\varDelta^2$ terms which, strictly speaking, are of higher order than the chiral order of the corresponding
Lagrangian.

In the following
we explicitly show how to calculate the loop diagrams:

Diagram Fig.~1(c) reads
\begin{equation}
A^{\alpha\mu,3}_{(c)}=-\sqrt{\frac{2}{3}}\left[\frac{h_A g_A^2}{(8\pi f_\pi)^2}\frac{1}{M_\Delta}\right]
i G^{\alpha\mu}_{(c)}
\end{equation}
with
\begin{equation}
iG^{\alpha\mu}_{(c)}=(2\pi\mu)^{4-D}\int\frac{d^D
k}{i\pi^2}\frac{ p'_b\gamma^{b\alpha c}
k_c\left[\slashpm-\slashk+M_N\right]\gamma^\mu\gamma_5\left[\slashp-\slashk+M_N\right]\slashk\gamma_5 
}{[k^2-m_\pi^2+i\epsilon][(p-k)^2-M_N^2+i\epsilon][(p'-k)^2-M_N^2+i\epsilon]},
\end{equation}
where $\mu$, the renormalization scale, is set to be $M_\Delta$.

Diagram Fig.~1(e) reads
\begin{equation}
A^{\alpha\mu,3}_{(e)}=\frac{5}{6}\sqrt{\frac{2}{3}}\left[\frac{g_A h_A H_A}{(8\pi
f_\pi)^2}\frac{1}{M^2_\Delta}\right]
iG^{\alpha\mu}_{(e)},
\end{equation}
with
\begin{equation}
iG^{\alpha\mu}_{(e)}=(2\pi\mu)^{4-D}\int\frac{d^D
k}{i\pi^2}\frac{ i \epsilon^{\alpha a b c}
p'_b k_c S_{ad}(p'-k)\gamma^{ed\mu}(p'-k)_e(\slashp-\slashk+M_N)\slashk\gamma_5
}{[k^2-m_\pi^2+i\epsilon][(p-k)^2-M_N^2+i\epsilon]}.
\end{equation}

Diagram Fig.~1(g) reads
 \begin{equation}
 A^{\alpha\mu,3}_{(g)}=\frac{1}{6}\sqrt{\frac{2}{3}}\left[\frac{h^3_A}{(8\pi f_\pi)^2} \frac{1}{M^3_\Delta}\right]
 iG^{\alpha\mu}_{(g)}
 \end{equation}
 with
 \begin{equation}
 iG^{\alpha\mu}_{(g)}=(2\pi\mu)^{4-D}\int\frac{d^D
k}{i\pi^2}\frac{
p'_a\gamma^{a\alpha b} k_b (\slashpm-\slashk-M_N)\gamma^{c\beta\mu}(p-k)_c
S_{\beta\gamma}(p-k)\gamma^{d\gamma e}(p-k)_d
k_e }{[k^2-m_\pi^2+i\epsilon][(p'-k)^2-M_N^2+i\epsilon]}.
 \end{equation}

 Diagram Fig.~1(i) reads
 \begin{equation}
 A^{\alpha\mu,3}_{(i)}=\frac{5}{9}\sqrt{\frac{2}{3}}\left[\frac{h_A H^2_A}{(8\pi
 f_\pi)^2}\frac{1}{M^4_\Delta}\right] iG^{\alpha\mu}_{(i)}
 \end{equation}
 with
 \begin{eqnarray}
 iG^{\alpha\mu}_{(i)}&=& (2\pi\mu)^{4-D}\int\frac{d^D
k}{i\pi^2} \\
&\times&
\frac{i\epsilon^{\alpha
a\rho\sigma}(p'-k)_\rho k_\sigma
S_{ab}(p'-k)\gamma^{lbm}\gamma^\mu\gamma^5\gamma_{lcn}(p'-k)_m(p-k)^n S^{cd}(p-k)\gamma_{fdg}(p-k)^f
k^g
}{[k^2-m_\pi^2+i\epsilon]}.\nonumber
 \end{eqnarray}
In the above equations, $S^{\mu\nu}(p)$  is the spin-3/2 propagator defined in
Eq.~(\ref{eq:3halfprop}).
 Since the couplings we used 
are spin-3/2 gauge  symmetric, our results do not depend on the specific value of
the gauge fixing parameter.

These loop functions are quite complicated, particularly the ones
including $\Delta$ internal lines. In practice, we adopt the conventional Feynman parametrization method (see Appendix B) and calculate these loop functions numerically.  
The manipulation of the Dirac algebra has been performed independently  with FORM~\cite{Vermaseren:2000nd} and
FeynCalc~\cite{Mertig:1990an}. The resulting Feynman parameter integrals are listed in Appendix C. Whenever possible,
the numerical results have been checked using the FF library~\cite{vanOldenborgh:1989wn} through
the LoopTools interface~\cite{Hahn:1998yk}.

 The one-loop results contain only four different
 Lorentz structures (due to the constraints $\bar{\Delta}_\alpha\gamma^\alpha=0$ and
$\bar{\Delta}_\alpha p'^\alpha=0$), i.e., $\gamma^\mu q^\alpha$, $q^\alpha p'^{\mu}
 $, $g^{\alpha\mu}$, and $q^\alpha q^\mu$. In accordance with the 
Adler formulation of Eq.~(\ref{Adlerff}), we can identify the corresponding Lorentz structures
and group the results as
 \begin{eqnarray}\label{loop_to_ff}
A^{\alpha\mu,3}_{(c)}+A^{\alpha\mu,3}_{(e)}+A^{\alpha\mu,3}_{(g)}+A^{\alpha\mu,3}_{(i)}
&=&\sqrt{\frac{2}{3}}\bigg[
g_3(q^2)\left(g^{\alpha\mu}\gamma\cdot q-q^\alpha\gamma^\mu\right)
 +g_4(q^2)\left(q\cdot p' g^{\alpha\mu}-q^\alpha p'^{\mu}\right)\nonumber\\
&&\hspace{4cm}+g_5(q^2)
 g^{\alpha\mu}+g_6(q^2)q^\alpha q^\mu\bigg].
 \end{eqnarray}
 
It is interesting to note that these loop results depend only on known masses and couplings:
$m_\pi$, $M_N$, $M_\Delta$, $f_\pi$, $g_A$, $h_A$, and $H_A$.
Here, we adopt the following values: $m_\pi=0.139$\,GeV, $M_N=0.939$\,GeV, $M_\Delta=1.232$\,GeV, 
$f_\pi=0.0924$\,GeV, $g_A=1.267$, $h_A=2.85$, and $H_A=(9/5)g_A$. The value of $H_A$ is
obtained from large $N_c$ relations and its uncertainty is discussed below. In other words, 
the $q^2$ dependence of the
loop functions are genuine predictions of the present work, in contrast with the $\delta^{(2)}$ and $\delta^{(3)}$ tree level diagrams, which
contain basically unknown low energy constants: $d_1$, $d_2$, $d_3$, $d_4$, 
$f_1$, $f_2$, $f_3$, $f_4$, $f_5$, $f_6$, and $f_7$. Some of these LEC, ($d_3$, $d_4$, $f_5$, $f_7$), also appear in pion-nucleon scattering and could, in principle, be extracted from there~\cite{Fettes:2000bb}.

Apart from diagrams (a),  (c), (e), (g), and (i), the external axial source can also
couple to a pion and interact through it with the system. These are the so-called pion pole terms (diagrams (b), (d), (f), (h), and (j)) and are calculated below. 

The Lagrangian
responsible for the coupling of the external axial source
with the pion at second order is
\begin{equation}
\mathcal{L}^{(2)}=-f_\pi\partial_\mu\pi^a a^{a,\mu}.
\end{equation}
With this and the low-energy counter terms given above, we can easily write
down the pion-pole contributions:
\begin{eqnarray}
A^{\alpha\mu,3}_\mathrm{pion-pole}&=&\sqrt{\frac{2}{3}}\frac{q^\alpha
q^\mu}{q^2-m^2_\pi}\bigg[\frac{h_A}{2}+ d_3 \gamma\cdot q +\frac{d_4}{M_\Delta}p'\cdot q -f_2 q^2-\frac{f_5}{M_\Delta}p'\cdot q\gamma\cdot q \nonumber\\
&&\hspace{6cm}-\frac{f_7}{M^2_\Delta}(p'\cdot q)^2-(g_5+g_6
q^2)\bigg]
\end{eqnarray}
with $g_5$ and $g_6$ the loop functions calculated above.

\section{Results and discussions}
In this section, 
we present our results for the form factors 
in terms of the LEC and the loop functions $g_3$, $g_4$, $g_5$, and $g_6$ (Table~\ref{table:avf}).
It should be mentioned that the Partially Conserved Vector Current (PCAC) relation
\begin{equation}
 C^A_5+\frac{C^A_6}{M^2_N}q^2|_{m_\pi\rightarrow0}=0
\end{equation}
holds up to every order in our $\chi$PT study, which can be easily checked from
Table~\ref{table:avf}.

\begin{table}[t]
      \renewcommand{\arraystretch}{1.5}
     \setlength{\tabcolsep}{0.4cm}
     \centering
     \caption{The axial transition form factors in relativistic baryon chiral perturbation theory;
 $d_1$, $d_2$, $d_3$, $d_4$ are order 2 LEC (in units of GeV$^{-1}$) while $f_1$, $f_2$, $f_3$, $f_4$, $f_5$, $f_6$, $f_7$ are order 3 LEC (in units of GeV$^{-2}$); $g_3(q^2)$, $g_4(q^2)$, $g_5(q^2)$, and $g_6(q^2)$ are the one-loop contributions as defined by Eq.~(\ref{loop_to_ff}).  \label{table:avf}}
     \begin{tabular}{c|ccc}
     \hline\hline 
      FF &  $\delta^{(1)}$ & $\delta^{(2)}$ & $\delta^{(3)}$\\
     \hline
$ -\sqrt{\frac{3}{2}}\frac{C^A_3(q^2)}{M_N}$& 0&$-d_2$ &$f_3\varDelta+g_3(q^2)$\\
 $-\sqrt{\frac{3}{2}}\frac{C^A_4(q^2)}{M^2_N}$&0 &$-d_1/M_\Delta$&$ (f_4+f_6)\varDelta/M_\Delta+g_4(q^2)$\\
$- \sqrt{\frac{3}{2}}C^A_5(q^2)$&$-\frac{h_A}{2}$&$-(d_3+d_4)\varDelta$&$(f_5+f_7) \varDelta^2+(f_1+f_2)q^2 +g_5(q^2)$\\
 $-\sqrt{\frac{3}{2}}\frac{C^A_6(q^2)}{M^2_N}$&$\frac{h_A/2}{q^2-m^2_\pi}$&$\frac{(d_3+d_4) \varDelta}{q^2-m^2_\pi}$&$-f_1+g_6(q^2)+\frac{-(f_5+f_7)\varDelta^2-f_2
 q^2-(g_5(q^2)+g_6(q^2) q^2)}{q^2-m^2_\pi}$\\
    \hline\hline
    \end{tabular} 
       \end{table}

As mentioned above,  the one-loop results are free of unknown couplings, but the LEC are basically not known.
Since these LEC always appear in particular combinations, we can introduce
$\tilde{d_1}=d_1-(f_4+f_6)\varDelta$, $\tilde{d}_2=d_2-f_3\varDelta$, and $\tilde{d}_3=d_3+d_4-(f_5+f_7)\varDelta$ and treat them as free parameters.
 Therefore, effectively, we have five unknown constants: $\tilde{d}_1$, $\tilde{d}_2$,
$\tilde{d}_3$, $f_1$, and $f_2$.

From Table \ref{table:avf}, we can conclude that
\begin{enumerate}
 \item[(a)] At order $\delta^{(1)}$, $C^A_3=0$, $C^A_4=0$, and $C^A_5=\sqrt{\frac{2}{3}}\frac{h_A}{2}\approx1.16$  with $h_A=2.85$ from Ref.~\cite{Pascalutsa:2005nd}, which is
 determined from the $\Delta$-resonance width, $\Gamma_\Delta=0.115\,\mathrm{GeV}$. 
Furthermore, $C^A_6$ is related to $C^A_5$ through the
pion-pole mechanism, i.e.,
  \begin{equation}\label{eq:c5ac6a}
 C^A_6=C^A_5 \frac{M^2_N}{m^2_\pi-q^2}.
 \end{equation}

\item[(b)] At order $\delta^{(2)}$,  $C^A_3$, $C^A_4$,  and $C^A_5$ receive a finite constant contribution. The above relation, Eq.~(\ref{eq:c5ac6a}),  between $C^A_5$ and $C^A_6$ still holds.

\item[(c)] At order $\delta^{(3)}$, the LEC give constant contributions to all form factors,  and $q^2$ dependent contributions to $C^A_5$ and $C^A_6$. The one-loop diagrams start  at this order.
\end{enumerate}

\begin{figure}[t]
\includegraphics[scale=1.5]{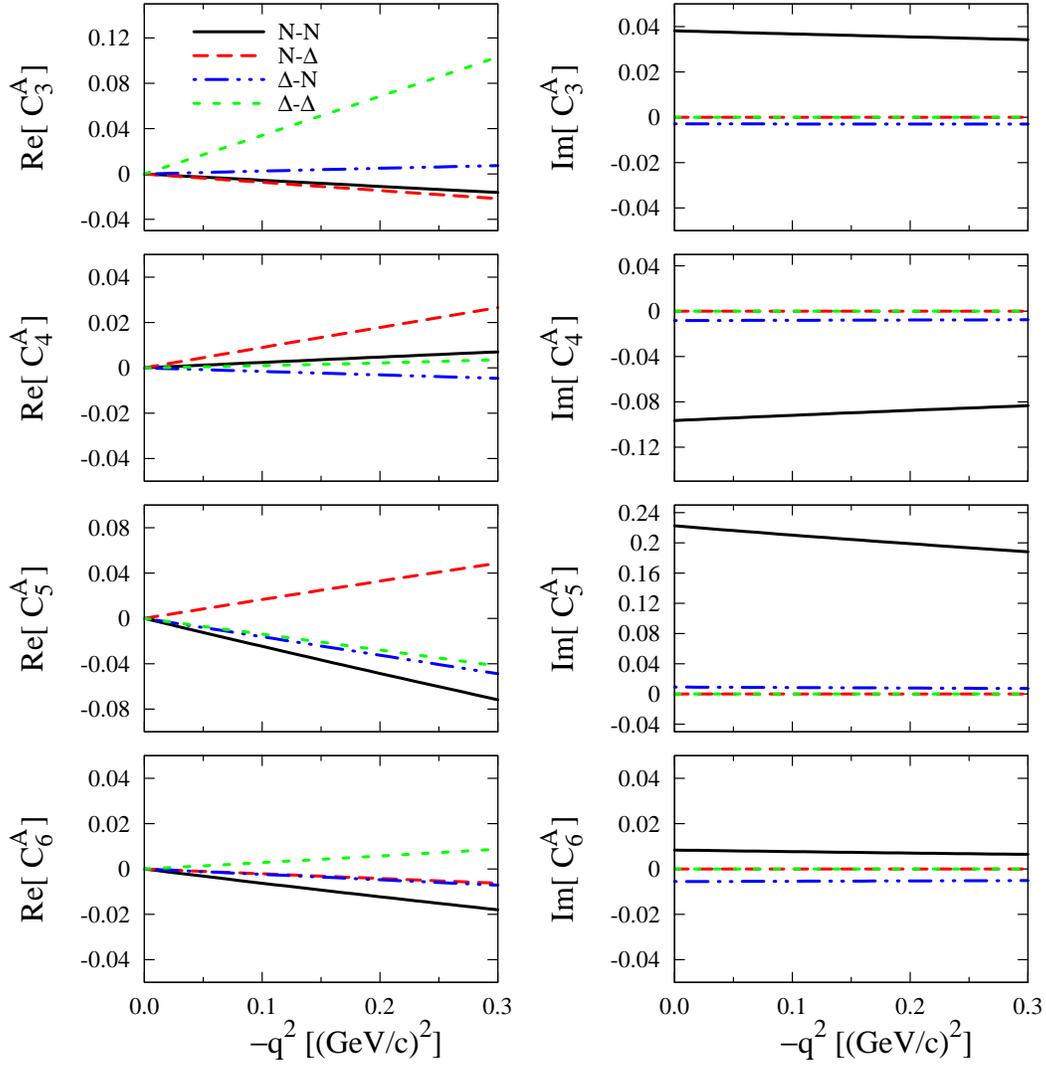}
\caption{(Color online) One-loop contributions to the form factors $C^A_3$, $C^A_4$, $C^A_5$, and $C^A_6$.
The pion-pole diagrams, which only contribute to $C^A_6$, have not been included. The
$N$-$N$, $N$-$\Delta$, $\Delta$-$N$, $\Delta$-$\Delta$ labels denote the contributions of
diagrams with nucleon-nucleon, nucleon-Delta, Delta-nucleon, Delta-Delta internal lines.
\label{fig:loops1}}
\end{figure}

Before presenting the loop results we specify our regularization procedure due to the complications with the power counting mentioned in Section II.A. The loops are regularized in the $\overline{MS}$ scheme, subtracting in addition the real part of the contribution to the form factors at $q^2$=0. Since there is no counter terms linear in $q^2$ at $\delta^{(2)}$, this procedure guarantees to recover the power counting in all form factors. 

We show in Fig.~\ref{fig:loops1} the one-loop contributions to the form factors $C^A_3$, $C^A_4$, $C^A_5$, and $C^A_6$ (except
the pion-pole diagrams which only contribute to $C^A_6$).
One can see that only diagrams \textit{c}, \textit{d} ($N$-$N$) and \textit{g}, \textit{h}
($\Delta$-$N$) from Fig.~1 contribute to the imaginary part of the form factors, with $N$-$N$ being dominant.
One also finds that
$C^A_4$ and $C^A_6$ receive relatively small corrections from the one-loop calculation, whereas $C^A_3$ gets a relatively large
one coming from the $\Delta$-$\Delta$ diagrams (diagrams \textit{i}, \textit{j}). This seemingly large $q^2$ dependence, however, suffers from the uncertainty
related to the $\pi\Delta\Delta$ coupling $H_A$ because the $\Delta$-$\Delta$ loop contribution is proportional to $H^2_A$.

\begin{figure}[t]
\includegraphics[scale=0.6]{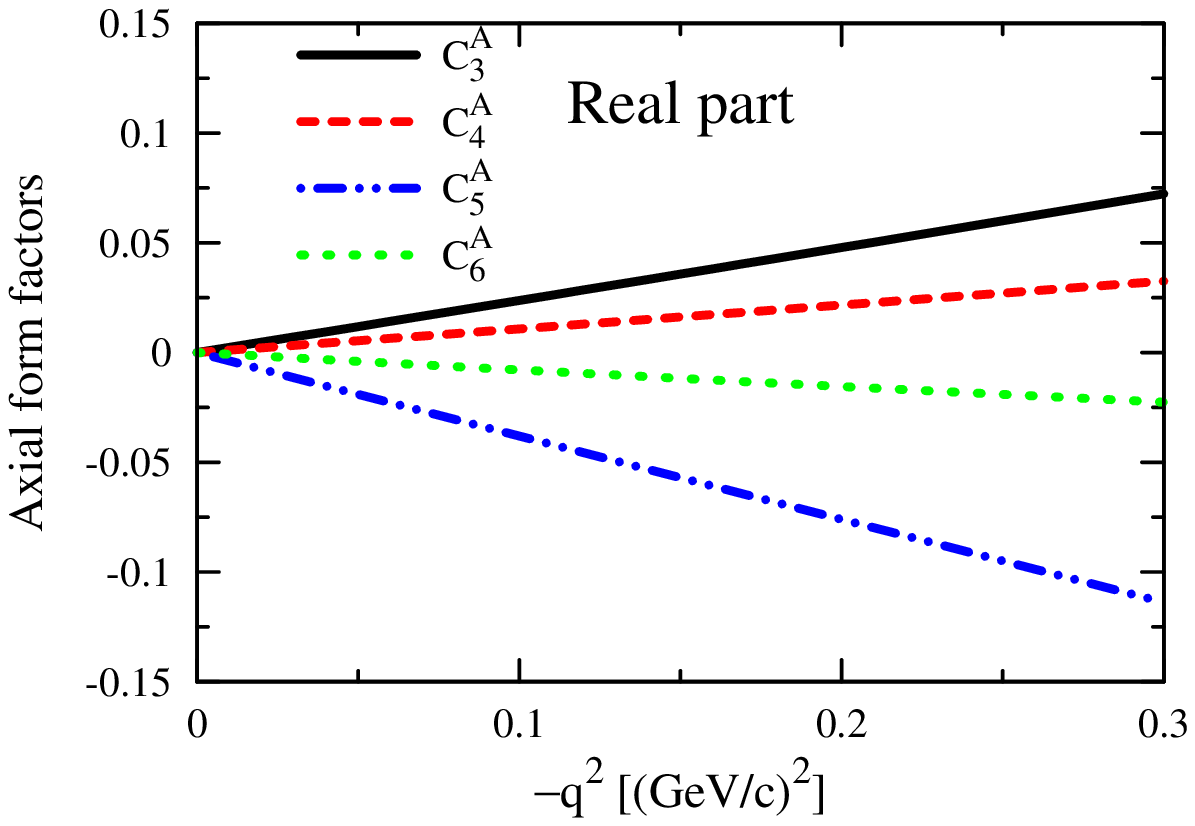}
\includegraphics[scale=0.6]{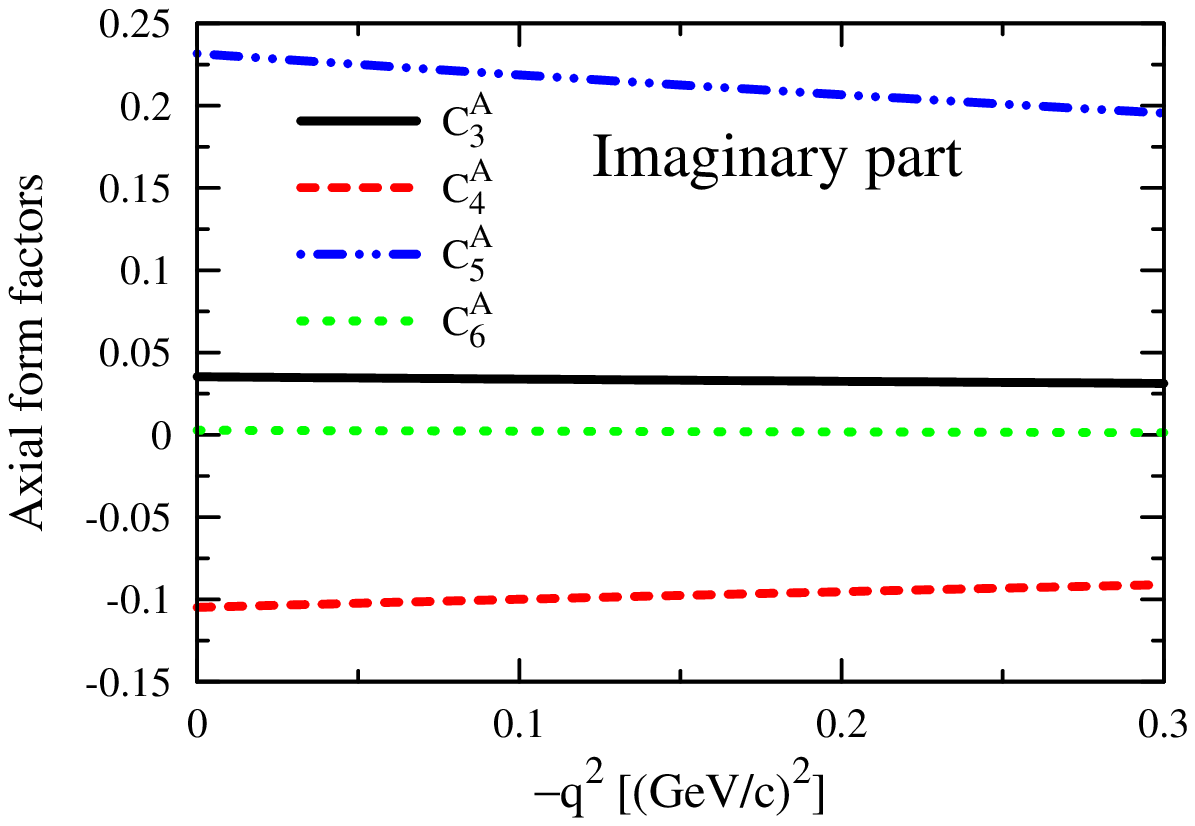}
\caption{(Color online) One-loop contributions to the form factors $C^A_3$, $C^A_4$, $C^A_5$, and $C^A_6$.
The pion-pole diagrams, which only contribute to $C^A_6$, have not been included
\label{fig:loops2}}
\end{figure}

In Fig.~\ref{fig:loops2}, the loop contributions from all diagrams to each form factor are added. Clearly, one can see that $C^A_5$ has
the largest imaginary part; $C^A_4$ the second; next is the $C^A_3$, and $C^A_6$ receives the smallest contribution.

Without the one-loop contributions, $C^A_6$ can be easily separated into a non-pole part and a pion-pole part, i.e.,
\begin{equation}
 C^A_6=-\tilde{g}_{\pi N\Delta} M^2_N\sqrt{\frac{2}{3}}\left[\frac{1}{q^2-m^2_\pi}+\frac{1}{6}r^2_A\right]
\end{equation}
with
\begin{equation}
 \tilde{g}_{\pi N\Delta}=\frac{h_A}{2}+\tilde{d}_3\varDelta -f_2
m^2_\pi,
\end{equation}
\begin{equation}
 r_A^2=-\frac{6}{\tilde{g}_{\pi N\Delta}}(f_1+f_2)\approx6\frac{d}{dq^2}\log(C^A_5)|_{q^2=0}.
\end{equation}
This is equivalent to the HB$\chi$PT result of Ref.~\cite{Zhu:2002kh}
\begin{equation}
 r^2_A=-\frac{6}{\Lambda_x^2}\frac{1}{g_{\pi N\Delta}}\left[ \frac{\tilde{b_3}+\tilde{b_8}}{2}\frac{\Lambda_x}{M_N}+c_2\right],
\end{equation}
with the correspondence $\tilde{g}_{\pi N\Delta}=g_{\pi N\Delta}$ and $(f_1+f_2)=\frac{1}{\Lambda_x^2}\left[ \frac{\tilde{b_3}+\tilde{b_8}}{2}\frac{\Lambda_x}{M_N}+c_2\right]$.

\section{Comparison with other approaches}
\subsection{Phenomenological fits}

Bubble chamber neutrino data have been used  to extract information about the
axial $N\rightarrow\Delta$ form factors \cite{Kitagaki:1990vs,AlvarezRuso:1998hi,Lalakulich:2005cs,Hernandez:2007qq}.
However, there are some important limitations. First, the cross section is basically
dominated by the $C_5^A$ form factor and shows very little sensitivity to $C_{3,4,6}^A$.
Second, the statistics is quite low and, furthermore, the two available data
sets from BNL~\cite{Kitagaki:1990vs} and ANL~\cite{Radecky:1981fn} are clearly different. 
Finally, it is difficult to disentangle
the $\Delta$ from other background pion production processes \cite{Sato:2003rq,Hernandez:2007qq}.
Therefore, all these works make some additional assumptions. A set of them often found
in the literature\footnote{This choice originates from the analysis of Refs. \cite{Bijtebier:1970ku,
LlewellynSmith:1971zm} of Adler's results
obtained using dispersion relations \cite{Adler:1968tw}. }
is: $C^A_3$=0, $C^A_4=-\frac{1}{4}C^A_5$, and
$C^A_6$ is related to $C^A_5$ through Eq.~(\ref{eq:c5ac6a}).
In this way only $C^A_5$ is fitted to the experiment. As an example, we can take Kitagaki et
al.~\cite{Kitagaki:1990vs} where it has the following functional form:
\begin{equation}
C_{5}^A(q^2) = {{C_5^A(0)\left[ 1-{{a_5 q^2}\over{b_5-q^2}} \right] }
{\left( 1- {{q^2}\over{M_A^2}}\right)^{-2}}}
\end{equation}
with $C_5^A(0)=1.2$, $a_5=-1.21$, $b_5=2$ GeV$^2$, and $M_A$ is fitted to data yielding
$M_A=1.28^{+0.08}_{-0.10}$~GeV. We will refer to this
set of form factors as Kitagaki-Adler (KA) form factors.

As we have shown above, there are 5 independent parameters in the $\delta$ expansion scheme
up to chiral order 3: $\tilde{d}_1$, $\tilde{d}_2$, $\tilde{d}_3$, $f_1$ and $f_2$.
We fix them in such a way that the real part of our form factors reproduces many of the features of the KA ones. 
To obtain $C^A_3=0$, we set $\tilde{d}_2=0$; therefore its contribution comes only from loop calculations
 which are of chiral order 3. Strictly speaking , Eq.~(\ref{eq:c5ac6a}) is not fulfilled at order $\delta^{(3)}$ 
but, if one neglects the small loop contributions, it can be satisfied by taking $f_1=0$.
Correspondingly, the  relation $C^A_4(0)=-\frac{1}{4}C^A_5(0)$ fixes
 $\tilde{d}_1$;  $C^A_5(0)=1.2$ fixes $\tilde{d}_3$.
The only LEC left, $f_2$, is then fixed to reproduce  $\frac{\partial C^A_5}{\partial q^2}$
at $q^2=0$. 

\begin{figure}[t]
\includegraphics[scale=1.0]{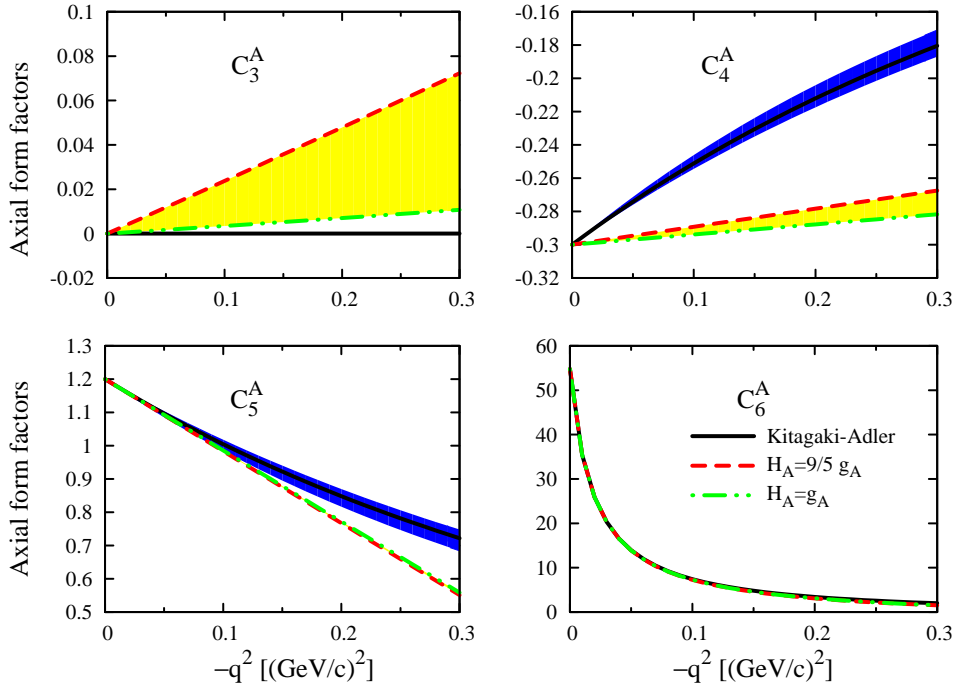}
\caption{(Color online) Comparison with the Kitagaki-Adler form factors. The dark shadowed area indicates 
the uncertainty of $M_A=1.28^{+0.08}_{-0.10}$\,GeV as determined in Ref.~\cite{Kitagaki:1990vs}. The light shadowed area
indicates the sensitivity of the results to the $\pi\Delta\Delta$ coupling $H_A$, which covers $H_A=(9/5)g_A$ to
$H_A=g_A$.
\label{fig:Adler}}
\end{figure}

The results obtained this way are shown in Fig.~\ref{fig:Adler}, with the following
parameter values $\tilde{d}_1=-0.514$ GeV$^{-1}$, $\tilde{d}_2=0$, $\tilde{d}_3=0.153$ GeV$^{-1}$,
$f_1=0$, and $f_2=-2.184$ GeV$^{-2}$. 
One can clearly see that
the calculated $C^A_5$ and $C^A_6$ are in good agreement with the KA form factors. 
On the other hand, the $q^2$ dependence of $C^A_4$ is much weaker that the one assumed in KA,
$C^A_4=-\frac{C^A_5}{4}$, and
we cannot accommodate their results at order $\delta^{(3)}$.
For $C^A_3$, the $q^2$ dependence is also very weak (compared to
$C^A_5$).  In Fig.~\ref{fig:Adler}, the dark shadowed area indicates 
a modification of $M_A$ within its uncertainties as given in Ref.~\cite{Kitagaki:1990vs}.
As we mentioned above, the $C^A_3$ dependence on $q^2$ is rather sensitive to the coupling constant
$H_A$. This can be easily seen from the light shadowed area in the upper-left panel of
Fig.~\ref{fig:Adler}, which covers the region of $g_A\le H_A\le (9/5) g_A$. The form factors, $C^A_{4,5,6}$, on the other hand, are less sensitive to the value of $H_A$.

A word of caution is in place about the comparison of $C^A_3$ and $C^A_4$ with the KA form factors. In 
$\chi$PT, the leading order counter terms linear in $q^2$ contributing to $C^A_3$ and $C^A_4$  appear at
chiral order 4.  A fair comparison with the phenomenological
fits (particularly the $q^2$ dependence) should, in principle, be done at order 4. 
However, the $\delta^{(3)}$ $\chi$PT results might give us a clue on the magnitude of the $q^2$ dependence of $C^A_3$ and $C^A_4$. Furthermore, if we believe in the phenomenological assumption, or the results of other approaches, the difference between
the third order $\chi$PT results and the results of other approaches might help us estimate
the value of the corresponding fourth order LEC. Indeed, the upper panels of Fig.~\ref{fig:Adler} indicate that small $\delta^{(4)}$ corrections $h_i q^2$ with {\it natural} values for the LEC $h_i$ can reproduce the slope assumed for $C^A_3$ and $C^A_4$ by the KA ansatz.

We also notice that a  recent analysis~\cite{Hernandez:2007qq} obtained a smaller value for $C^A_5(0)$ 
by including non $\Delta$ contributions and fitting to the low invariant mass
ANL data. In the present $\chi$PT study, we do have the higher-order contributions, 
$\tilde{d}_3$, which could alter
$C^A_5(0)$ within such a range; however, the same LEC appear in
pion-nucleon scattering processes. A combined analysis is mandatory to determine whether one can accommodate the small $C^A_5(0)$ obtained in, for instance, Ref.~\cite{Hernandez:2007qq}. This is left for future studies.
\subsection{Quark models}

There have been many studies of the $N\rightarrow\Delta$ axial  transition form factors in various quark models,
both relativistic and non-relativistic. For a brief review of quark model studies, we refer the readers 
to Refs.~\cite{Liu:1995bu,BarquillaCano:2007yk}. Compared 
to dynamical model studies, a feature of most quark model calculations is that the obtained form factors are real due to
time-reversal symmetry, while in dynamical models, like our $\chi$PT study, these form factors are
in general complex due to the opening of the pion-nucleon channel.

Quark model results are in fact quite scattered. Taking, for instance,
the models discussed in Ref.~\cite{BarquillaCano:2007yk}, we observed that the prediction of $C^A_5(0)$ runs from 0.81 to 1.53, $C^A_4(0)$ runs from $-0.66$ to 0.14 and $C^A_3(0)$ runs from 0 to 0.05. These models also obtain the non-pole part of $C^A_6$ whose value at 
$q^2=0$ ranges from $-0.72$ to 1.13. 
We could use these results to extract our constants although the large differences between them do not allow to reach solid conclusions about their values. From $C^A_5(0)$  one gets $\tilde{d}_3$, and from its slope $\partial C^A_5/\partial q^2$ at $q^2=0$, $(f_1+f_2)$. This fixes the non-pole part of $C^A_6(0)$ (neglecting the one-loop corrections) since 
\begin{equation}
C^{A\mathrm{(non-pole)}}_6(0) \approx \sqrt{\frac{2}{3}}M^2_N(f_1+f_2) \,, 
\end{equation}
which is nothing but a direct consequence of PCAC. Using the quark model calculation of Ref.~\cite{BarquillaCano:2007yk} for $ C^A_5$ we obtain $C^{A\mathrm{(non-pole)}}_6(0) \approx -2$. This value is almost a factor three larger in magnitude than the one obtained directly from that model in spite of the fact that it implements PCAC at the quark level by introducing one- and two-body axial exchange currents. 

\begin{figure}[t]
\includegraphics[scale=0.62]{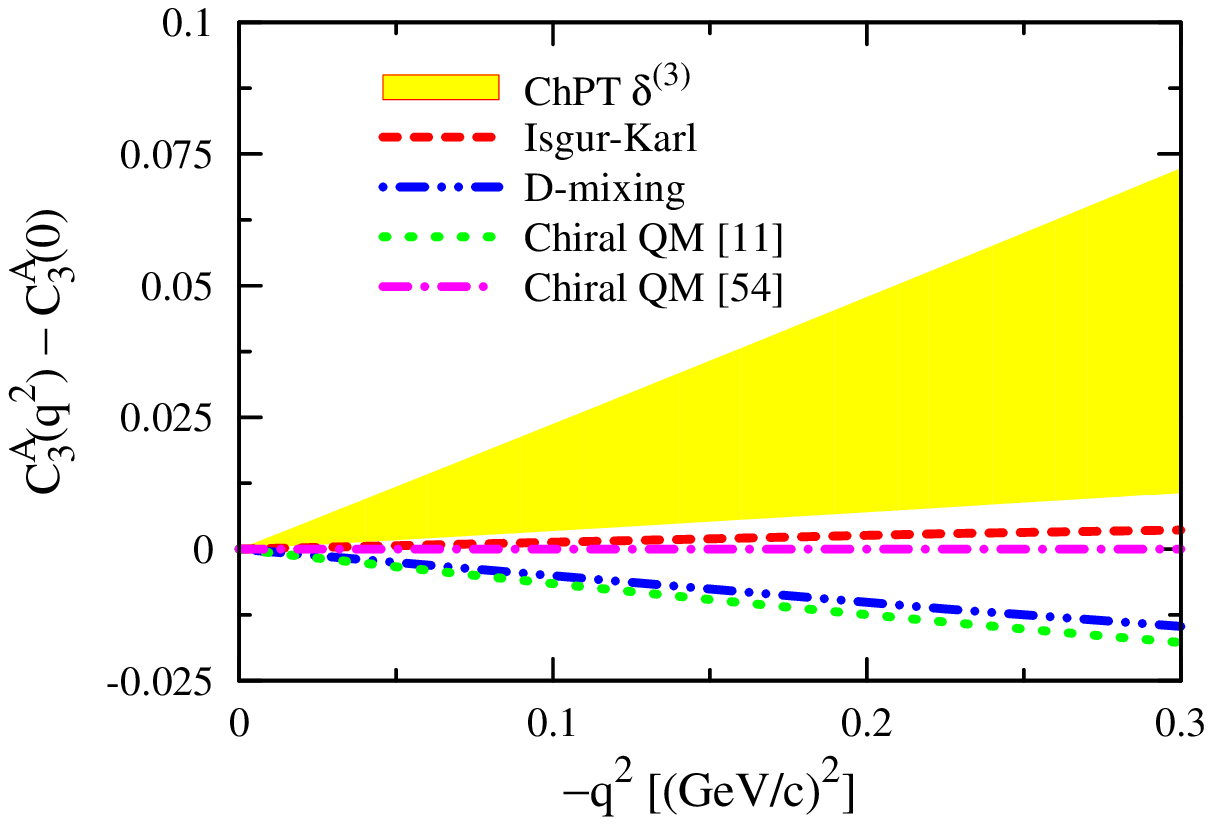}\hfill
\includegraphics[scale=0.62]{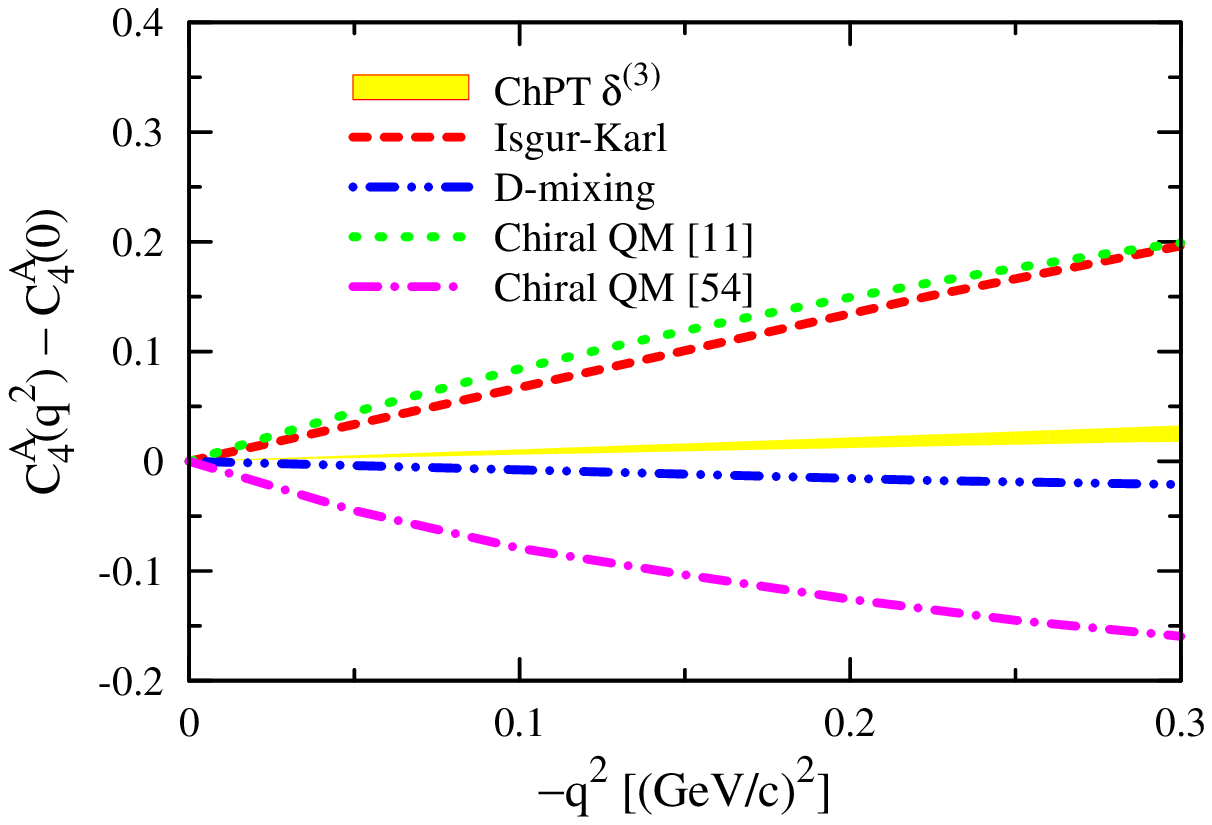}
\caption{(Color online) Comparison with the non relativistic  Isgur-Karl and D-mixing  quark model results of Ref.~\cite{Liu:1995bu},  and those of the chiral quark models of Refs.~\cite{BarquillaCano:2007yk} and ~\cite{Golli:2002wy}.
\label{fig:Quark}}
\end{figure}

Analogously, we can use quark model results for $C^A_3$ and $C^A_4$ at $q^2=0$ to obtain $\tilde{d}_1$ and $\tilde{d}_2$. The smallness of  $C^A_3(0)$ predicted by all calculations points towards a $\tilde{d}_2$ close to zero, in agreement with the phenomenological assumption. The situation is much more uncertain 
with $\tilde{d}_1$, both in sign and magnitude. 
In Fig.~\ref{fig:Quark}, the $q^2$ dependence of  the real parts of $C^A_3$ and $C^A_4$ in our calculation, which at order $\delta^{(3)}$ is dictated by the loops, is compared to several quark models. As in the case of the KA form factors discussed above, we can expect from this comparison that next order terms linear in $q^2$ with small ({\it natural}) values of the LEC are sufficient to eliminate the discrepancies in the low $q^2$ behavior with any of these quark models.

\subsection{Lattice QCD results}
Recently, the $N\rightarrow\Delta$ axial  transition form factors have been studied in lattice QCD~
\cite{Alexandrou:2006mc,Alexandrou:2007zz}. Some major conclusions are
(i) $C^A_3$ and $C^A_4$ are suppressed compared to $C^A_5$ and $C^A_6$, and (ii) $C^A_5$ can be described by a dipole ansatz
$C^A_5(0)/(1+Q^2/M^2_A)^2$ but with a smaller $C^A_5(0)$ and a larger $M_A$ ($\gtrsim1.5$ GeV), compared to
the Kitagaki-Adler form factors. These results should be taken with caution because of
the still relatively large pion mass ($\ge350$ MeV) used in the study.

In principle, $\chi$PT is the perfect tool to extrapolate the lattice QCD results to the physical region. Meanwhile, one can
also fix the unknown couplings to the lattice QCD results. Due to the regularization method we used and the fact that the lattice data points are still scarce, we will leave this subject to the future.

\section{Summary and conclusions}
We have studied the axial $N\rightarrow\Delta$ transition
form factors up to one-loop order in relativistic 
baryon chiral perturbation theory with the $\delta$ expansion scheme.
The adopted Lagrangians including the $\Delta(1232)$
are {\it consistent}, i.e., spin-3/2 gauge symmetric, which automatically
decouples unphysical spin-1/2 fields. Consequently, our results do not depend on
the specific value of the gauge-fixing parameter that is present in the most general spin-3/2
propagator, and avoid various problems related to inconsistent couplings.

The form factor $C^A_5$  exhibits the richest structure in our study. It receives contributions starting at chiral order 1, at which we find that $C^A_5(0)=\sqrt{\frac{2}{3}}\frac{h_A}{2}\approx1.16$ for $h_A=2.85$. At higher orders, this value is modified by low energy constants that are unknown but which also appear in pion-nucleon scattering. At chiral order 3, this form factor gets $q^2$ dependent contributions, some of them complex. Actually, we find that  $C^A_5$ has the largest imaginary part among the four form factors. 
We also obtain that, up to chiral order 2, $C^A_6=C^A_5\frac{M^2_N}{m^2_\pi-q^2}$. At order 3, 
$C^A_6$ has a non-pole contribution whose value at $q^2=0$ is related to the slope of $C^A_5$  at $q^2=0$. Assuming natural values for the LEC, this non-pole part is small compared to the dominant pion-pole mechanism.

Both  $C^A_3$ and $C^A_4$ start at chiral order 2 and get their $q^2$ dependence at order 3 from the loops.
For $C^A_3$, we find a  small $q^2$ dependence, which is quite sensitive to the $\pi\Delta\Delta$ coupling constant, $H_A$. On the other hand, its imaginary part, coming mainly from
the $N$-$N$ internal diagram, is finite ($\sim0.03$ at $q^2=0$) and has a mild $q^2$ dependence. This suggests that $C^A_3$ is small (compared to $C^A_{4,5,6}$) but not necessarily zero. The $C^A_4$ dependence on $q^2$  is also found to be rather mild at order $\delta^{(3)}$. 

We have compared our results with a phenomenological set of form factors used in the analysis of neutrino-induced pion production data and also with different quark model calculations. 
They could be used to extract the low energy constants but the scarcity of data and the large differences between quark model results make it difficult to come to solid conclusions. In the case of $C^A_3$ and specially $C^A_4$, the comparison should, in principle, be done at order 4 where corresponding leading order counter terms linear in  $q^2$ appear.  Nevertheless we can say that reasonable agreement with all these approaches can be obtained with natural values of the LEC. 

Future experiments with electron and neutrino beams, combined with the analysis of pion-nucleon scattering data, can shed more light on these form factors. The extrapolation of lattice QCD results to the physical region should also be pursued.

\section{Acknowledgments}
We thank Mauro Napsuciale, Stefan Scherer, Wolfram Weise,
 and in particular Massimiliano Procura and Vladimir Pascalutsa for useful
discussions. We are also grateful to Eliecer Hernandez for providing us with the results of several quark model calculations.  L. S. Geng acknowledges financial
support from the Ministerio de Educacion y Ciencia in the Program 
``Estancias de doctores y tecnologos extranjeros''. J. Martin Camalich acknowledges  the same institution for a FPU fellowship. This work was partially supported by the  MEC 
contract  FIS2006-03438, the Generalitat Valenciana ACOMP07/302,
and the EU Integrated Infrastructure
Initiative Hadron Physics Project contract RII3-CT-2004-506078.

\textbf{Note added in proof:} After submitting this paper, a new preprint~\cite{Procura:2008ze} appeared that studies
the $N\rightarrow\Delta$ axial form factors up to one-loop order in HBChPT using  the small scale expansion scheme. 
Within this framework, there is  no $q^2$ dependence coming from the loop-functions. This supports  the smooth 
$q^2$ dependences  found in  the present work. Namely,  
the $q^2$ dependence of the loops in our relativistic framework is  counted as of higher-order in HBChPT.
 \bibliographystyle{apsrev}
 \bibliography{axialff}

\section{Appendix}
\subsection{Isospin transition matrices and antisymmetric Gamma matrix products}
The isospin 1/2 to 3/2 and 3/2 to 3/2 transition matrices $T^a$ and
$\mathcal{T}^a$ appearing in the $N\Delta$ and $\Delta\Delta$ Lagrangians are given
by:
\begin{equation}
\setlength{\arraycolsep}{0.3 cm}
 T^1=\frac{1}{\sqrt{6}}\left(
\begin{array}{cccc}
-\sqrt{3}&0&1&0\\
0&-1&0&\sqrt{3}\end{array} \right),
\end{equation}
\begin{equation}
\setlength{\arraycolsep}{0.3 cm}
 T^2=\frac{-i}{\sqrt{6}}\left(
\begin{array}{cccc}
\sqrt{3}&0&1&0\\
0&1&0&\sqrt{3}\end{array} \right),
\end{equation}
\begin{equation}
\setlength{\arraycolsep}{0.3 cm}
 T^3=\sqrt{\frac{2}{3}}\left(
\begin{array}{cccc}
0&1&0&0\\
0&0&1&0\end{array} \right).
\end{equation}
\begin{equation}
\setlength{\arraycolsep}{0.3 cm}
 \mathcal{T}^1=\frac{2}{3}\left(
\begin{array}{cccc}
0&\sqrt{3}/2&0&0\\
\sqrt{3}/2&0&1&0\\
0&1&0&\sqrt{3}/2\\
0&0&\sqrt{3}/2&0\end{array} \right),
\end{equation}
\begin{equation}
\setlength{\arraycolsep}{0.3 cm}
 \mathcal{T}^2=\frac{2i}{3}\left(
\begin{array}{cccc}
0&-\sqrt{3}/2&0&0\\
\sqrt{3}/2&0&-1&0\\
0&1&0&-\sqrt{3}/2\\
0&0&\sqrt{3}/2&0\end{array} \right),
\end{equation}
\begin{equation}
\setlength{\arraycolsep}{0.3 cm}
 \mathcal{T}^3=\left(
\begin{array}{cccc}
1&0&0&0\\
0&1/3&0&0\\
0&0&-1/3&0\\
0&0&0&-1\end{array} \right).
\end{equation}

The totally antisymmetric Gamma matrix products appearing in the {\it consistent}
$N\Delta$ and $\Delta\Delta$ Lagrangians are defined as:
\begin{equation}
 \gamma^{\mu\nu}=\frac{1}{2}[\gamma^\mu,\gamma^\nu],
\end{equation}
\begin{equation}
 \gamma^{\mu\nu\rho}=\frac{1}{2}\{\gamma^{\mu\nu},\gamma^{\rho}\}=-i\varepsilon^{\mu\nu\rho\sigma}\gamma_\sigma\gamma_5,
\end{equation}
\begin{equation}
 \gamma^{\mu\nu\rho\sigma}=\frac{1}{2}[\gamma^{\mu\nu\rho},\gamma^\sigma]=i\varepsilon^{\mu\nu\rho\sigma}\gamma_5
\end{equation}
with the following conventions: $g^{\mu\nu}=\mathrm{diag}(1,-1,-1,-1)$,  $\varepsilon_{0,1,2,3}=-\varepsilon^{0,1,2,3}=1$,
$\gamma_5=i\gamma_0\gamma_1\gamma_2\gamma_3$.

\subsection{Loop functions}
In the calculation of the loop diagrams, we have used the following
$d$-dimensional integrals in Minkowski space:
\begin{equation}
\int d^d k\frac{k^{\alpha_1}\ldots k^{\alpha_{2n}}}{(\mathcal{M}^2-k^2)^\lambda}
=i\pi^{d/2}\frac{\Gamma(\lambda-n+\epsilon-2)}{2^n\Gamma(\lambda)}\frac{(-1)^n g^{\alpha_1
\ldots\alpha_{2n}}_s}{(\mathcal{M}^2)^{\lambda-n+\epsilon-2}}
\end{equation}
with $g^{\alpha_1 \ldots\alpha_{2n}}_s=g^{\alpha_1\alpha_2}\ldots g^{\alpha_{2n-1}\alpha_{2n}}+\ldots$
a combination symmetrical with respect to the permutation of any pair of indices (with $(2n-1)!!$
terms in the sum)~\cite{Smirnov:2006ry}.

The $\mathcal{M}^2$ that appear in the calculation of the $N$-$N$, $N$-$\Delta$, $\Delta$-$N$, and $\Delta$-$\Delta$ internal diagrams are, respectively,
\begin{equation}
 \mathcal{M}_{NN}^2=x\,m_\pi^2-x(1-x-y) M_\Delta^2+(1-x-x y) M_N^2-y(1-x-y)q^2-i\epsilon,
\end{equation}
\begin{equation}
 \mathcal{M}_{N\Delta}^2=x\,m_\pi^2+(1-x)(1-x-y)M_\Delta^2+y(1-x)M_N^2-y(1-x-y)q^2-i\epsilon,
\end{equation}
\begin{equation}
 \mathcal{M}_{\Delta N}^2=x\,m_\pi^2 + (x^2 + x y - x + y) M_\Delta^2 + (1 - x - y - x y) M_N^2 - 
 y (1 - x - y) q^2-i\epsilon,
\end{equation}
\begin{equation}
 \mathcal{M}_{\Delta \Delta}^2=x\,m_\pi^2 + (1 - 2 x + x^2 + x y) M_\Delta^2 - x y M_N^2 - y (1 - x - y) q^2-i\epsilon,
\end{equation}
where $x$ and $y$ are Feynman parameters.
\subsection{Feynman parameterization integrals}

We present below the loop integrals, diagrams (c), (e), (g) and (i)  of Fig. 1, cast in the Feynman parameterization. We use the following notation: $\widetilde{C}_i^{(XY)}$ is the $\overline{MS}$-regularized contribution of the loop to $C_i^A$ with $X$ and $Y$ being the baryons in the internal line (in this order), $\bar{\mathcal{M}}^2_{XY}$=$\mathcal{M}^2_{XY}/M_\Delta^2$, $r=M_N/M_\Delta$, $\mu_\pi=m_\pi/M_\Delta$, and $\bar{Q}^{2n}=Q^{2n}/M_\Delta^{2n}$ (with $Q^2=-q^2$).

 The couplings are contained in the constants ${\cal C}_{XY}$:	
\begin{eqnarray}
{\cal C}_{NN}&=&2\,\sqrt{\frac{2}{3}}\;\frac{g_A^2 h_A M_{\Delta }^2}{128 f_{\pi }^2 \pi ^2}\;\;;\;\;{\cal C}_{N\Delta}=\frac{5}{3}\,\sqrt{\frac{2}{3}}\;\frac{g_A h_A H_A M_{\Delta }^2}{192 f_{\pi }^2 \pi ^2}\;\;;\nonumber\\
{\cal C}_{\Delta N}&=&\frac{1}{3}\,\sqrt{\frac{2}{3}}\;\frac{h_A^3 M_{\Delta }^2}{192 f_{\pi }^2 \pi ^2}\,\;\;;\;\;{\cal C}_{\Delta\Delta}=\frac{10}{9}\,\sqrt{\frac{2}{3}}\;\frac{h_A H_A^2 M_{\Delta }^2}{576 f_{\pi }^2 \pi ^2}\,.\nonumber
\end{eqnarray}

Then, the expressions of the loop functions are:
\begin{eqnarray}
\widetilde{C}_3^{(NN)}&=&-\frac{{\cal C}_{NN}}{24} r (14 r+3)+r\,{\cal C}_{NN}\,\int_0^1 dx\int_0^{1-x} dy \Big\{\Big[y \left(\left(-3 y+x \left(y^2+2 y-1\right)+1\right) \right.
   r^3\nonumber\\
&+& \left. \left((y-1) x^2+\left(y^2-y-1\right) x-3 y+2\right) r^2-x
   (y+3) (x+y-1) r+x \left(-x^2-2 y x\right.\right.\nonumber\\
&+&\left.\left. x-y^2+y\right)+\bar{Q}^2 (x+y-1)
   \left((y-1) (x+y)+r \left(y^2+2 y-1\right)\right)\right)\Big]\frac{1}{\bar{\mathcal{M}}^2
   _{NN}}\nonumber\\
&-&  \Big[2 y (2 y-1)+x (4 y-1)+r \left(4 y^2+5 y-1\right)\Big]
   \log \left(\bar{\mathcal{M}}^2 _{NN}\right)\Big\}\,,\nonumber\\
\widetilde{C}_4^{(NN)}&=&\frac{r^2\,{\cal C}_{NN}}{6}-2\,r^2\,{\cal C}_{NN}\,\int_0^1 dx\int_0^{1-x} dy\,y\,\Big\{\Big[\left(-3 y+x \left(y^2+y-2\right)+2\right)
   r^2-2 x (x \nonumber\\
&+& y-1) r-y (x+y-1) \left((1-y)
   \bar{Q}^2+x\right)\Big]\frac{1}{\bar{\mathcal{M}}^2 _{NN}}-2(2 y-1) \log \left(\bar{\mathcal{M}}_{NN}^2\right)\Big\}\,,\nonumber\\
\widetilde{C}_5^{(NN)}&=&-\frac{1-r}{r}\widetilde{C}_3^{(NN)}-\frac{1-r^2-\bar{Q}^2}{2\,r^2} \widetilde{C}_4^{(NN)}-2\, r\, {\cal C}_{NN}\,\int_0^1 dx\int_0^{1-x} dy\,((1-r)y +r+x)\times\nonumber\\
&\times&\log \left(\bar{\mathcal{M}}^2 _{NN}\right)\,,\nonumber\\
\widetilde{C}_6^{(NN)}&=&\frac{r^2\,{\cal C}_{NN}}{6}-2\,r^2\,{\cal C}_{NN}\,\int_0^1 dx\int_0^{1-x} dy\,y\,\Big\{\Big[y((x (y-1)-2 y+1) r^2+2 (x+y-1) r\nonumber\\
&-& x (x+y-1)+\bar{Q}^2
   (y-1) (x+y-1))\Big]\frac{1}{\bar{\mathcal{M}}^2_{NN}}-2(2 y-1) \log \left(\bar{\mathcal{M}}^2 _{NN}\right)\Big\}\,.\nonumber\\
\widetilde{C}_3^{(N\Delta)}&=&\frac{{\cal C}_{N\Delta}}{288} r \left(-39 \mu _{\pi }{}^2+155 r^2+27 \bar{Q}^2+96 r-69\right)-r\, {\cal C}_{N\Delta}\,\int_0^1 dx\int_0^{1-x} dy\,\times\nonumber\\
&\times&\Big\{  y \Big[x (x (y-1)-2 y+1) y r^4+y \left(-2 x^2-3 y
   x+x+1\right) r^3+\left((1-2 y) x^3+(3-2 y) y
   x^2\right.\nonumber\\
&+&\left. \left(y^2-1\right) x+y\right) r^2+x \left(2 x^2+5 y x+3
   y^2-y-2\right) r+x (x+1) \left(x^2+(2 y-1) x\right.\nonumber\\
&+& \left. (y-1)
   y\right)+\bar{Q}^4 (y-1) y \left(x^2+(2 y-1) x+ (y-1) y\right)+\bar{Q}^2
   \left((1-2 y) x^3+\left(2 (y-1) r^2\right.\right.\nonumber \\
&-& \left. \left.2 r-4 y+3\right) y
   x^2+\left(2 \left(r^2-1\right) y^3+\left(-5 r^2-5 r+1\right)
   y^2+\left(2 r^2+r+2\right) y-1\right) x\right.\nonumber\\
&+&\left. y \left(\left(y-2
   y^2\right) r^2+\left(-3 y^2+2 y+1\right)
   r-(y-1)^2\right)\right)\Big]\frac{1}{\bar{\mathcal{M}}^2 _{N\Delta}}-\Big[(4-33 y) x^2+\left(33 \left(r^2\right.\right.\nonumber\\
&+& \left.\left. \bar{Q}^2-1\right)
   y^2-\left(21 r^2+32 r+21 \bar{Q}^2-5\right) y+4\right) x+y
   \left((13-32 y) r^2+(8-48 y) r-16 y\right.\nonumber\\
&+& \left.\bar{Q}^2 \left(33 y^2-38
   y+13\right)+8\right)\Big] \frac{\log \left(\bar{\mathcal{M}}^2
   _{N\Delta}\right)}{4}+13 (4 y-1)\frac{\bar{\mathcal{M}}^2
   _{N\Delta}\, \log \left(\bar{\mathcal{M}}^2
   _{N\Delta}\right)}{4}\Big\}\,,\nonumber
\end{eqnarray}
\begin{eqnarray}
\widetilde{C}_4^{(N\Delta)}&=&-\frac{r^2\,{\cal C}_{N\Delta}}{240} \left(-45 \mu _{\pi }{}^2+27 r^2+33 \bar{Q}^2-152
   r-55\right)+r^2\,{\cal C}_{N\Delta}\,\int_0^1 dx\int_0^{1-x} dy\,\times\nonumber\\
 &\times&\Big\{y\Big[(3 x ( x (y-1)-2 y+1) y r^4-y \left((y+3) x^2+5 y
   x-3\right) r^3+\left((2-5 y) x^3+(9-5 y) y x^2\right.\nonumber\\
&+& \left.\left(5
   y^2-y-2\right) x+y\right) r^2+x \left((y+4) x^2+y (y+8) x+5
   y^2-y-4\right) r+3 \bar{Q}^4 (y\nonumber\\
&-& 1) y \left(x^2+(2 y-1) x+(y-1)
   y\right)+x \left(2 x^3+4 y x^2+\left(2 y^2+y-2\right) x+(y-1)
   y\right)\nonumber\\
&-& \bar{Q}^2 \left((5 y-2) x^3+y \left(-6 (y-1) r^2+(y+3) r+10
   y-7\right) x^2+\left(\left(-6 r^2+r+5\right) y^3\right.\right.\nonumber\\
&+&\left.\left.\left(15 r^2+7
   r-4\right) y^2-3 \left(2 r^2+1\right) y+2\right) x+y \left(3 y
   (2 y-1) r^2+\left(5 y^2-2 y-3\right)
   r\right.\right.\nonumber\\
&+&\left. \left.(y-1)^2\right)\right)\Big]\frac{1}{2 \bar{\mathcal{M}}^2 _{N\Delta}}+\Big[(19 y-2) x^2+\left(\left(-24 r^2+5 r-24
   \bar{Q}^2+19\right) y^2+\left(15 r^2+14 r\right.\right.\nonumber\\
&+& \left.\left.15 \bar{Q}^2-5\right) y-2\right)
   x+y \left(3 (8 y-3) r^2+4 (5 y+1) r+4 y-3 \bar{Q}^2 \left(8 y^2-9
   y+3\right)-2\right)\Big]\times\nonumber\\
&\times& \frac{\log \left(\bar{\mathcal{M}}^2
   _{N\Delta}\right)}{2}+(4 y-1) \frac{9\bar{\mathcal{M}}^2 _{N\Delta}\,\log \left(\bar{\mathcal{M}}^2
   _{N\Delta}\right)}{2}\Big\}\,, \nonumber\\
\widetilde{C}_5^{(N\Delta)}&=&-\frac{1-r}{r}\,\widetilde{C}_3^{(N\Delta)}-\frac{1-r^2-\bar{Q}^2}{2\,r^2}\, \widetilde{C}_4^{(N\Delta)}+\frac{{\cal C}_{N\Delta}}{1440} \left(5\left(-27 \bar{Q}^2+9 (43-3 r) r+158\right) \mu _{\pi }{}^2\right.\nonumber\\
&+& \left.99
   \bar{Q}^4+5 \bar{Q}^2 (r (36 r-191)-57)+r (r (r (81
   r+385)+135)+5175)+290\right)\nonumber\\
&-& {\cal C}_{N\Delta}\,\int_0^1 dx\int_0^{1-x} dy\,\Big\{\Big[\left((r-1)^2+\bar{Q}^2\right) y \left((y-1) y (x+y-1) (x+y)
   \bar{Q}^4\left((1\right.\right.\nonumber\\
&-&\left.\left.2 y) x^3+(2 r (r (y-1)-1)-4 y+3) y x^2+\left(y
   \left((y (2 y-5)+2) r^2-5 y r+r-2 y^2+y\right.\right.\right.\right.\nonumber\\
&+& \left.\left.\left.\left.2\right)-1\right) x+y
   \left(-(r+1) (2 r+1) y^2+(r (r+2)+2) y+r-1\right)\right)
   \bar{Q}^2+(r-1) (r+1)^2 \times\right.\nonumber\\
&\times& \left.(r (x-2)-x-1) x y^2+x (r (r+2)+x)
   \left(x^2-1\right)-(r+1) \left(-x r^3-r^2+2 (r-1) x^3\right.\right.\nonumber\\
&+&\left.\left.\left(r
   \left(r^2+r-4\right)-1\right) x^2+x\right) y\right)\Big]\frac{3}{4 \bar{\mathcal{M}}^2
   _{N\Delta}}-\Big[x^3+4 r^2 x^2+2 r x^2+6 x^2+6 r^2 x-2 r x-x\nonumber\\
&-& 4 r^2+(r-1)^2 (r+1) (24 r (x-1)-23 x-12) y^2+8 r-(r-1) \left(-9
   r^3+4 r^2+2 r\right.\nonumber\\
&+&\left.(25 r-22) x^2+3 (r (r (5 r+6)-5)+2) x+6\right)
   y+3 \bar{Q}^4 y (y (8 y-9)+x (8 y-5)+3)\nonumber\\
&-&\bar{Q}^2 \left((25 y-4) x^2+(y
   (2 y+r (6 r (5-8 y)+47 y+3)+13)-6) x+y \left(-3 (y (8 y-17)\right.\right.\nonumber\\
&+& \left.\left.6)
   r^2+(y (47 y-8)+13) r+(41-23 y) y-22\right)+4\right)\Big]\frac{\log \left(\bar{\mathcal{M}}^2 _{N\Delta}\right)}{4}-\Big[4 (x-8 y+5)\nonumber\\
&+&9 \bar{Q}^2 (1-4 y)+r
   (9 r (1-4 y)+68 y+7)\Big] \frac{\bar{\mathcal{M}}^2
   _{N\Delta}\,\log \left(\bar{\mathcal{M}}^2
   _{N\Delta}\right)}{4}\Big\}\,,\nonumber\\
\widetilde{C}_6^{(N\Delta)}&=&-\frac{r^2\,{\cal C}_{N\Delta}}{360} (199 r+3)+r^2\,{\cal C}_{N\Delta}\,\int_0^1 dx\int_0^{1-x}\,y\, dy\, \Big\{y\Big[((r-2) r+x) (x-1)^2+\left(r^2+r\right.\nonumber\\
&-&\left.2\right) (r-x) y
   (x-1)+(r-1)^2 (r+1) x y^2-\bar{Q}^2 (x+y-1) (x (y-1)-y (r+(r-1)
   y\nonumber\\
&+& 2)+1)\Big]\frac{1}{2 \bar{\mathcal{M}}^2 _{N\Delta}}+\Big[(x+y-1) (5 y-2)-5  r y
   (y+1)\Big] \frac{\log \left(\bar{\mathcal{M}}^2_{N\Delta}\right)}{2}\Big\}\,.\nonumber
\end{eqnarray}
\newpage
\begin{eqnarray}
 \widetilde{C}_3^{(\Delta N)}&=&\frac{r\,{\cal C}_{\Delta N}}{1440} \left(5 (73 r+95) \mu _{\pi }{}^2+2 \bar{Q}^2 (11 r+54)+r (r (345
   r+467)+175)+359\right)\nonumber\\
&+& r\,{\cal C}_{\Delta N}\,\int_0^1 dx\int_0^{1-x} dy\,\Big\{\Big[\left((r+1)^2+\bar{Q}^2\right) y (x+y-1) \left(x (y-1)^2
   r^3-\left(y+x \left(y^2\right.\right.\right.\nonumber\\
&+& \left.\left.\left.x (y-1)\right)-1\right) r^2+\left(\bar{Q}^2
   (x+y-1) (y-1)^2+y+x (-y x+x-(y-2) y+1)\right.\right.\nonumber\\
&-& \left.\left. 1\right) r+(x+y-1)
   \left(x^2+\left(-y \bar{Q}^2+\bar{Q}^2+y+1\right) x-\bar{Q}^2 y^2\right)\right)\Big]\frac{1}{4
   \bar{\mathcal{M}}^2_{\Delta N}}-\Big[\left(2 x \left(4 y^2\right.\right.\nonumber\\
&-& \left.\left.6 y+1\right)+(y-1) (y (5
   y-4)+1)\right) r^3+\left(5 y^3-(x+17) y^2+(9-2 x (3 x+5)) y\right.\nonumber\\
&+& \left.2
   x^2+x-1\right) r^2+\left((3-13 y) x^2+2 (2-9 y) y x+y (12-y (5
   y+7))\right) r+(x-5 y\nonumber\\
&+& 1) (x+y)^2-2 (r+x-2 y)+2 \bar{Q}^2 (x+y-1)
   \left((4 r-3) y^2-(6 r+3 x+2) y+r+x\right)\Big]\times\nonumber\\
&\times& \frac{\log\left(\bar{\mathcal{M}}^2_{\Delta N}\right)}{4}- (x+r (2-3 y)+y+1)\bar{\mathcal{M}}^2 _{\Delta N} \log \left(\bar{\mathcal{M}}^2 _{\Delta N}\right)\Big\}\,,\nonumber\\
\widetilde{C}_4^{(\Delta N)}&=&-\frac{r^2\,{\cal C}_{\Delta N}}{720} \left(120 \mu _{\pi }{}^2+8 \bar{Q}^2+r (175r+107)+190\right)+r^2\,{\cal C}_{\Delta N}\,\int_0^1 dx\int_0^{1-x} dy\times\nonumber\\
&\times&\Big\{\Big[y (x+y-1) \left(-x y (x+y-1) r^4+(y+x (x+(x+2) y-1)-1)
   r^3+\left(x \left(2 y^2+4 x y\right.\right.\right.\nonumber\\
&+& \left.\left.\left.(x-2) x-2\right)+\bar{Q}^2 \left(y
   \left(-y^2-4 x y+y-2 (x-2) x+1\right)-1\right)\right)
   r^2+\left(-y+\bar{Q}^2 \left((x \right.\right.\right.\nonumber\\
&+&\left.\left.\left. 2) y^2+ x (x+4) y-3 y+(x-4) x+1\right)-x
   (2 y+x (x+y+3))+1\right) r+x-\bar{Q}^4 y (x\right.\nonumber\\
&+& \left.y-1) (x+2 y-2)-x (y+(x+y)
   (2 x+y))+\bar{Q}^2 \left(x^3+(5 y-3) x^2+(y (5 y-4)\right.\right.\nonumber\\
&+&\left.\left.1) x+y
   \left(y^2+y-2\right)\right)\right)\Big]\frac{1}{2 \bar{\mathcal{M}}^2 _{\Delta
   N}}-\Big[x^3+\left(\left(17-6 \bar{Q}^2\right) y-3\right)
   x^2+\left(y \left(2 \bar{Q}^2 (9-11 y)\right.\right.\nonumber\\
&+&\left.\left.21 y-8\right)+1\right) x+r
   \left((5 x+8) y^2+x (5 x+14) y-7 y+(x-4) x+1\right)+r^2 \left(y
   \left(-6 x^2\right.\right.\nonumber\\
&+&\left.\left.4 (3-4 y) x+(3-5 y) y+3\right)-1\right)+y \left(5 y
   (y+1)-4 \bar{Q}^2 (y (4 y-7)+3)-6\right)\Big]\frac{\log \left(\bar{\mathcal{M}}^2
   _{\Delta N}\right)}{2}\nonumber\\
&-& 2  (x+6 y-2)\bar{\mathcal{M}}^2
   _{\Delta N} \log \left(\bar{\mathcal{M}}^2
   _{\Delta N}\right)\nonumber\Big\}\,,\nonumber\\
\widetilde{C}_5^{(\Delta N)}&=&-\frac{1-r}{r}\,\widetilde{C}_3^{(\Delta N)}-\frac{1-r^2-\bar{Q}^2}{2\,r^2}\, \widetilde{C}_4^{(\Delta N)}-\frac{{\cal C}_{\Delta N}}{1440}\left(5 \left(5 \bar{Q}^2+r (26 r+63)+5\right) \mu _{\pi }{}^2+8 \bar{Q}^4\right.\nonumber\\
&+&\left.\bar{Q}^2\left(42 r^2+52 r+15\right)+r (r (r (193
   r+289)+114)+223)+7\right)-{\cal C}_{\Delta N}\,\int_0^1 dx\int_0^{1-x} dy\times\nonumber\\
&\times&\Big\{\Big[\left(y^2-\left(x^2+2\right) y+1\right)
   r^4+\left(y+x \left(y^2+(x-2) y+x+2\right)-1\right)
   r^3+\left(x^3-2 \bar{Q}^2 (x\right.\nonumber\\
&-&\left.1) y x-\left(2 \bar{Q}^2 x+x+2\right) y^2+2
   y\right) r^2-(x+y-1) \left(x^2-\left(\bar{Q}^2+\left(\bar{Q}^2-1\right)
   y+1\right) x\right.\nonumber\\
&-&\left.\bar{Q}^2 y^2\right) r+\left(\bar{Q}^2+1\right) (x+y-1)
   \left(\left(\bar{Q}^2+1\right) y+(x+y) \left(x-\bar{Q}^2
   y\right)\right)\Big]\frac{ \log \left(\bar{\mathcal{M}}^2_{\Delta
   N}\right)}{4}\nonumber\\
&+&\left(x+r ((r-1) x-y)+y+\bar{Q}^2
   (x+y-1)-1\right) \bar{\mathcal{M}}^2_{\Delta N}\log \left(\bar{\mathcal{M}}^2_{\Delta N}\right)\Big\}\,,\nonumber
\end{eqnarray}
\newpage
\begin{eqnarray}
 \widetilde{C}_6^{(\Delta N)}&=&-\frac{r^2\,\mathcal{C}_{\Delta N}}{720} \left(95 \mu _{\pi }{}^2+16 \bar{Q}^2+r (121
   r+91)+199\right)-r^2\,\mathcal{C}_{\Delta N}\,\int_0^1 dx\int_0^{1-x} dy\times\nonumber\\
&\times&\Big\{\Big[y (x+y-1) \left((y-1) (x y+y-1) r^4+x ((y-4) y+3) r^3+\left(2
   \left(x^2+x+y\right)\right.\right.\nonumber\\
&-&\left.\left.y \left(4 x^2+3 y x+2 y\right)+\bar{Q}^2 (y-1)
   \left(y^2+2 x y-1\right)\right) r^2+\left(\bar{Q}^2 \left((y-1)^2+x
   (y-3)\right) (y\nonumber\right.\right.\\
&-&\left.\left. 1)+y-x (x (y-5)+(y-4) y+1)-1\right) r+3 x^3+x^2
   \left(\bar{Q}^2 (2-4 y)+5 y-1\right)\right.\nonumber\\
&+&\left.(y-1) y \left((y-1) \bar{Q}^4-2 y
   \bar{Q}^2+1\right)+x \left(\left(\left(\bar{Q}^2-6\right) \bar{Q}^2+2\right)
   y^2-\left(\bar{Q}^2-6\right) \bar{Q}^2 y+y\right.\right.\nonumber\\
&-&\bar{Q}\left.\left.^2-1\right)\right)\Big]\frac{1}{2\bar{\mathcal{M}}^2
   _{\Delta N}}+\Big[\left(y \left((2 x (3-5 y)+y (3-5 y)+3) r^2+(x (18-5
   y)+(13-5 y) y\right.\right.\nonumber\\
&-&\left.\left.7) r+10 (x+y) (2 x+y)-6 (3 x+y)-2 \bar{Q}^2 (x+y-1) (5
   y-3)\right)-(r+x) (r+2 x\right.\nonumber\\
&-&\left.1)\right)\Big] \frac{\log \left(\bar{\mathcal{M}}^2
   _{\Delta N}\right)}{2}+2 (5 y-1)\bar{\mathcal{M}}^2
   _{\Delta N} \log \left(\bar{\mathcal{M}}^2
   _{\Delta N}\right)\Big\}\,. \nonumber\\
\widetilde{C}_3^{(\Delta\Delta)}&=&\frac{r\,{\cal C}_{\Delta\Delta}}{1440}\left(6620 \mu _{\pi }{}^4+\left(368 \bar{Q}^2+7 (483-320 r) r+22011\right) \mu _{\pi }{}^2+182 \bar{Q}^4+\bar{Q}^2 (2520\right.\nonumber\\
&-&\left.r (274 r+3))+r (r (r (648r-625)-7941)+3789)+32345\right)+r\,{\cal C}_{\Delta\Delta}\,\int_0^1 dx\int_0^{1-x} dy\times\nonumber\\
&\times&\Big\{\Big[\left((r+1)^2+\bar{Q}^2\right) y (x+y-1) \left(x^4+2 \left(\bar{Q}^2+(r-1) r-\left(r^2+\bar{Q}^2\right) y+y+1\right) x^3\right.\nonumber\\
&+&\left.\left((y-2) y \bar{Q}^4+\left(2
   (y-2) y r^2+(2 y-1) r-4 y^2+2 y-1\right) \bar{Q}^2-(r-2) r^2+\left(r^2-1\right)^2 y^2\right.\right.\nonumber\\
&-&\left.\left.2 r \left((r-1) r^2+2\right) y+2 (r+2 y-1)\right)
   x^2+\left(2 \bar{Q}^4 y (y-1)^2+(r-1) \left((2 (y-1) y-1) r^2\right.\right.\right.\nonumber\\
&-&\left.\left.\left.(y+1) r-2 y^2+y+3\right)+\bar{Q}^2 \left(2 r^2 y (y-1)^2+r (4 (y-1) y-1)+y (3-2 y
   (y+1))\right.\right.\right.\nonumber\\
&-&\left.\left.\left.3\right)\right) x+\bar{Q}^2 (y-1) y \left(\left(\bar{Q}^2 (y-1)-2\right) y+r (2 y-1)\right)\right)\Big]\frac{1}{4 \bar{\mathcal{M}}^2_{\Delta \Delta }}-\Big[\left(x y (x (18 y-19)\right.\nonumber\\
&+&\left.4 (y (3 y-7)+3)) r^4+\left(10 (2 x+1) y^3+(x (20 x-19)-12) y^2+2 (1-6 x) x y-6 y\right.\right.\nonumber\\
&-&\left.\left.2 x (x+2)+4\right)
   r^3+\left(2 (5-14 y) x^3+(4 (5-8 y) y+6) x^2+(y ((7-4 y) y+19)-20) x\right.\right.\nonumber\\
&-&\left.\left.11 y+5 y^2 (2 y-3)+8\right) r^2-\left(10 (2 x+1) y^3+\left(40
   x^2+x-4\right) y^2+x (4 x+1) (5 x-4) y\right.\right.\nonumber\\
&-&\left.\left.18 y-2 x ((x-1) x+6)+8\right) r+10 x^4+(7-10 y) y^2+23 y+x^3 (12 y+7)-2 x^2 (3 (y\right.\nonumber\\
&-&\left.3) y+19)+\bar{Q}^4 y
   (x+y-1) (3 y (6 y-5)+x (18 y-19)+5)+x \left(y \left(-8 y^2+y-27\right)\right.\right.\nonumber\\
&+&\left.\left.31\right)+\bar{Q}^2 \left(2 (5-14 y) x^3+(8 (1-6 y) y+2) x^2+y (77-6 y
   (2 y+7)) x-26 x-39 y\right.\right.\nonumber\\
&+&\left.\left.y^2 (8 (y-5) y+67)+r^2 y \left((36 y-38) x^2+4 (4 y (3 y-5)+9) x+(y-1)^2 (12 y-5)\right)\right.\right.\nonumber\\
&+&\left.\left.r \left(2 (2 y (5 y-3)-1)
   x^2+(5 y-2) (y (8 y-5)+2) x+(y-1) y (y (20 y-9)+11)+4\right)\right.\right.\nonumber\\
&+&\left.\left.12\right)-12\right)\Big] \frac{\log \left(\bar{\mathcal{M}}^2_{\Delta \Delta }\right)}{4}-\Big[\left(28 x+66 y+\bar{Q}^2 \left(-88 y^2+76 y+x (38-88 y)-28\right)\right.\nonumber\\
&-&\left.2 \left((5 y (3 y-5)+x (44 y-19)+7) r^2+(4 x (5
   y-1)+2 y (10 y-9)+7) r-(39 x-5 y)\right.\right.\times\nonumber\\
&\times& \left.\left.(x+y)\right)-67\right)\Big]\frac{\bar{\mathcal{M}}^2_{\Delta \Delta } \log \left(\bar{\mathcal{M}}^2_{\Delta \Delta }\right)}{4}-14\, \bar{\mathcal{M}}^4_{\Delta \Delta } \log \left(\bar{\mathcal{M}}^2_{\Delta \Delta }\right)\Big\}\,,\nonumber
\end{eqnarray}
\newpage
\begin{eqnarray}
\widetilde{C}_4^{(\Delta\Delta)}&=&-\frac{r^2\,{\cal C}_{\Delta\Delta}}{720} \left(2340 \mu _{\pi }{}^4+3 \left(72 \bar{Q}^2+(269-130 r) r+2261\right) \mu _{\pi }{}^2+60 \bar{Q}^4+\bar{Q}^2 (788-r (42 r\right.\nonumber\\
&+&\left.23))+r (r (r (90
   r+43)-1288)+1289)+8988\right)-r^2\,{\cal C}_{\Delta\Delta}\,\int_0^1 dx\int_0^{1-x} dy\times\nonumber\\
&\times&\Big\{\Big[y (x+y-1) \left(\left(5 \bar{Q}^2+r (5 r-2)-1\right) x^4+\left((4-8 y) \bar{Q}^4+(14 y+2 r (r (4-8 y)+y)\right.\right.\nonumber\\
&+&\left.\left.1) \bar{Q}^2-(r+1) (2 y+r (2 r (-5 y+r (4
   y-2)+2)-5)+3)\right) x^3+\left(3 (y-1) y r^6-y r^5\right.\right.\nonumber\\
&+&\left.\left.\left(-7 y^2+y+3\right) r^4+(3-6 y) r^3+(y (5 y+9)-4) r^2+7 y r+r+\bar{Q}^4 (r (9 r
   (y-1)-1)\nonumber\right.\right.\\
&-&\left.\left.19 y+11) y+3 \bar{Q}^6 (y-1) y-y (y+7)+\bar{Q}^2 \left(9 (y-1) y r^4-2 y r^3+(2 (6-13 y) y+3) r^2\right.\right.\right.\nonumber\\
&-&\left.\left.\left.6 y r+r+y (13 y+7)-7\right)-1\right)
   x^2+\left(3 y (y (2 y-3)+1) \bar{Q}^6+\left(y \left(6 (y (2 y-3)+1) r^2\right.\right.\right.\right.\nonumber\\
&+&\left.\left.\left.\left.2 \left(y^2+y-2\right) r+(5-14 y) y+7\right)-4\right) \bar{Q}^4+\left(2
   (r+1)^2 (r (3 r-5)+2) y^3+(10\nonumber\right.\right.\right.\\
&-&\left.\left.\left.r (r (r (9 r-8)+3)+10)) y^2+(r-1) \left(3 r^3-7 r^2+r+8\right) y-5 r^2+r-3\right) \bar{Q}^2+\left(r^2\right.\right.\right.\nonumber\\
&-&\left.\left.\left.1\right)
   \left(6 (y-1) y r^3+\left(-4 y^2+y-1\right) r^2+3 \left(-2 y^2+y+1\right) r+4 y^2-1\right)\right) x+\bar{Q}^2 (y-1)\times\right.\nonumber\\
&\times&\left. y \left(3 (y-1) y
   \bar{Q}^4+\left(3 (y-1) y r^2+(2 y (y+2)-3) r-y (3 y+5)+1\right) \bar{Q}^2+4 y+r (-6 y\right.\right.\nonumber\\
&+&\left.\left.r (-4 y+r (6 y-3)-1)+1)+1\right)\right)\Big]\frac{1}{2 \bar{\mathcal{M}}^2_{\Delta \Delta}}-
\Big[\left(3 x y \left(10 y^2+14 x y-16 y-9 x+6\right) r^4\right.\nonumber\\
&+&\left.y \left(6 (x+5) y^2+(33-4 x) x y-42 y-4 x (x+7)+12\right) r^3+\left(-2
   (23 x+10) y^3-6 \left(19 x^2\right.\right.\right.\nonumber\\
&+&\left.\left.\left.x-3\right) y^2+x \left(-68 x^2+46 x+31\right) y-12 y+2 x (x (7 x+5)-9)+4\right) r^2+\left(-6 (x+5) y^3\right.\right.\nonumber\\
&+&\left.\left.(x-1)
   (10 x-33) y^2+(x (x (16 x-29)+28)-1) y+4 x^2+6 x-4\right) r+4 (4 x+5) y^3\right.\nonumber\\
&+&\left.\left(46 x^2+44 x-17\right) y^2+2 (x-1) x (x (7 x+9)-9)+x (4 x
   (11 x+7)-49) y+13 y\right.\nonumber\\
&+&\left.3 \bar{Q}^4 y (x+y-1) (y (14 y-13)+x (14 y-9)+3)+\bar{Q}^2 \left(2 (7-34 y) x^3+2 \left(y \left(3 (14 y-9) r^2\right.\right.\right.\right.\nonumber\\
&-&\left.\left.\left.\left.2 (y+1) r-82
   y+35\right)+2\right) x^2+\left(y \left(6 (y (19 y-26)+9) r^2+(y (14 y+23)-20) r+4 (11\right.\right.\right.\right.\nonumber\\
&-&\left.\left.\left.\left.31 y) y+53\right)-26\right) x+y \left(3 (y-1) (2
   y-1) (5 y-3) r^2+(y (y (18 y+11)-40)+17) r\right.\right.\right.\nonumber\\
&+&\left.\left.\left.y (57-4 y (7 y+3))-29\right)+8\right)-4\right)\Big]\frac{\log \left(\bar{\mathcal{M}}^2_{\Delta \Delta }\right)}{2}+\Big[\left(-16 x-62 y+12 \bar{Q}^2 (y (13 y-12)\right.\nonumber\\
&+&\left.x (13 y-4)+3)+2 \left(3 (2 y (5 y-6)+x (26 y-8)+3) r^2-(10 y x+x-(3 y+5)
   (5 y-2)) r\right.\right.\nonumber\\
&-&\left.\left.(x+y) (53 x+25 y)\right)+39\right)\Big]\frac{\bar{\mathcal{M}}^2_{\Delta \Delta } \log \left(\bar{\mathcal{M}}^2_{\Delta \Delta }\right)}{2}-36 \bar{\mathcal{M}}^4_{\Delta \Delta } \log \left(\bar{\mathcal{M}}^2_{\Delta \Delta }\right)\Big\}\,,\nonumber\\
\widetilde{C}_5^{(\Delta\Delta)}&=&-\frac{1-r}{r}\,\widetilde{C}_3^{(\Delta\Delta)}-\frac{1-r^2-\bar{Q}^2}{2\,r^2}\, \widetilde{C}_4^{(\Delta\Delta)}+\frac{{\cal C}_{\Delta\Delta}}{1440}\left(20 \left(117 \bar{Q}^2+r (117 r-272)+117\right) \mu _{\pi }{}^4\right.\nonumber\\
&+&\left.\left(216 \bar{Q}^4+(2 (370-87 r) r+7263) \bar{Q}^2+r \left(r \left(-390 r^2+2012
   r+3773\right)-15462\right)\right.\right.\nonumber\\
&+&\left.\left.7047\right) \mu _{\pi }{}^2+60 \bar{Q}^6+\bar{Q}^4 (r (18 r-109)+930)+\bar{Q}^2 \left(r \left(48 r^3-717
   r+37\right)+11013\right)\right.\nonumber\\
&+&\left.r (r (r (r (r (90 r-347)-803)+5067)+3238)-21462)+10143\right)+{\cal C}_{\Delta\Delta}\,\int_0^1 dx\int_0^{1-x} dy\,\times\nonumber
\end{eqnarray}
\newpage
\begin{eqnarray}
&\times&\Big\{\Big[\left(r^4+2 \left(\bar{Q}^2-1\right) r^2+\left(\bar{Q}^2+1\right)^2\right) y (x+y-1) \left(x^4+\left(r^2+\bar{Q}^2-2 \left(r^2+\bar{Q}^2-1\right) y+1\right)
   x^3\right.\nonumber\\
&+&\left.\left((y-1) y r^4+\left(-2 y^2+2 \bar{Q}^2 (y-1) y+1\right) r^2-2 y r+r+y \left(\bar{Q}^2 \left(\bar{Q}^2 (y-1)-4 y+2\right)+y+3\right)\right.\right.\nonumber\\
&-&1\left.\left.\right)
   x^2+\left(y (y (2 y-3)+1) \bar{Q}^4+\left(y \left((y (2 y-3)+1) r^2+2 (y-1) r-y (2 y+1)+2\right)-1\right) \bar{Q}^2\right.\right.\nonumber\\
&+&\left.\left.(r-1) ((r+1) (2 r (y-1)-2 y+1)
   y+1)\right) x+\bar{Q}^2 (y-1) y \left(\left(\bar{Q}^2 (y-1)-2\right) y+r (2 y\right.\right.\nonumber\\
&-&\left.\left.1)\right)\right)\Big]\frac{3}{4 \bar{\mathcal{M}}^2_{\Delta \Delta }}-\Big[\left(11 x^3 r^4+8 x^2 r^4-9 x r^4-6 x^3 r^3+20 x r^3-4 r^3+10 x^4 r^2+9 x^3 r^2-12 x^2 r^2\right.\nonumber\\
&-&\left.x r^2+4 r^2-10 x^4 r-4 x^3 r+32 x^2
   r-12 x r+4 r+10 x^4+10 x^3+2 (r-1)^3 (r+1)^2 ((15 r\right.\nonumber\\
&+&\left.16) x+15) y^3-18 x^2+(r-1)^2 (r+1) \left(6 x (7 x-8) r^3+\left(30 x^2+x-42\right)
   r^2+3 \left(-20 x^2+x\right.\right.\right.\nonumber\\
&-&\left.\left.\left.4\right) r
-54 x^2+4 x+31\right) y^2+2 x-(r-1) \left(9 x (3 x-2) r^5+2 (x (10 x+9)-6) r^4+(x (x (52 x\right.\right.\nonumber\\
&-& \left.\left. 31)+7)
-6) r^3+\left(x \left(30 x^2-9 x-8\right)+9\right) r^2
+\left(x \left(-30 x^2+9 x+20\right)+6\right) r+4 (11-3 x) x^2\right.\right.\nonumber\\
&+&\left.\left.x+3\right) y
+3 \bar{Q}^6 y
   (x+y-1) (y (14 y-13)+x (14 y-9)+3)+\bar{Q}^4 \left((11-52 y) x^3+(y (-30 y \right.\right.\nonumber\\
&+&\left.\left.r (-12 y
+9 r (14 y-9)+7)-7)+2) x^2+\left(y \left(6 (11 y (3
   y-4)+15) r^2+((71-24 y) y-38) r\right.\right.\right.\right.\nonumber\\
&+&\left.\left.\left.\left.(y-1) (96 y
-113)\right)-17\right) x+y \left(18 (y-1) (r-2 r y)^2+y (y (74 y-191)+147)+2 r (11\right.\right.\right.\nonumber\\
&-&\left.\left.\left. 2 y (y (3
   y-16)+17))
-34\right)+4\right)+\bar{Q}^2 \left(3 y \left(3 (14 y-9) x^2+4 (y (12 y-17)+6) x+(y-1) (2 y\right.\right.\right.\nonumber\\
&-&\left.\left.\left.1) (5 y-3)\right) r^4+2 y \left((7
-12 y)
   x^2+((60-11 y) y-37) x+y (y (y+39)-54)+17\right) r^3+\left(2 (11\right.\right.\right.\nonumber\\
&-&\left.\left.\left.52 y) x^3+2 ((22-81 y) y
+5) x^2-(y (y (120 y+23)-76)+26) x+y (y ((33-62
   y) y+21)\right.\right.\right.\nonumber\\
&-&\left.\left.\left.14)+4\right) r^2+\left((22 y-6) x^3
+\left(42 y^2-34 y+2\right) x^2+(y (2 y (9 y+32)-79)+26) x-(y-1)\times \right.\right.\right.\nonumber\\
&\times&\left.\left.\left.(y (2 (y-45)
   y+51)-12)\right) r
+(x+y-1) \left(32 y^3-4 (13 x+27) y^2-x (74 x+41) y+22 y\right.\right.\right.\nonumber\\
&+&\left.\left.\left.x (2 x+5) (5 x+3)\right)\right)-4\right)\Big]
\frac{\log \left(\bar{\mathcal{M}}^2_{\Delta\Delta }\right)}{4}-\Big[\left(48 x r^4-18 r^4-26 x r^3+38 r^3+86 x^2 r^2+36 x r^2\right.\nonumber\\
&-&\left.21 r^2-82 x^2 r-16 x r+65 r
+86 x^2-10 (r-1)^2
   (r+1) (6 r+7) y^2+38 x-2 (r-1) (8 x\right.\nonumber\\
&+&\left.r (-28 x+r (32 x+6 r (13 x-6)+3)
+6)+67) y-12 \bar{Q}^4 (y (13 y-12)+x (13 y-4)+3)\right.\nonumber\\
&+&\left.\bar{Q}^2 \left(-6 \left(9
   (1-2 y)^2+4 x (13 y-4)\right) r^2
+2 \left(46 y^2-66 y+x (46 y-13)+22\right) r-226 y^2\right.\right.\nonumber\\
&+&\left.\left.86 x (x+1)-140 x y+278 y-75\right)-39\right)\Big]
\frac{\bar{\mathcal{M}}^2_{\Delta \Delta }\log\left(\bar{\mathcal{M}}^2_{\Delta \Delta }\right)\nonumber}{4}-2\left(9 \bar{Q}^2+r (9 r-8)+9\right)\times\nonumber\\
&\times& \bar{\mathcal{M}}^4_{\Delta \Delta } \log \left(\bar{\mathcal{M}}^2_{\Delta \Delta }\right)\Big\}\,,\nonumber
\end{eqnarray}
\newpage
\begin{eqnarray}
\widetilde{C}_6^{(\Delta\Delta)}&=&-\frac{r^2\,{\cal C}_{\Delta\Delta}}{720} \left(5 (36 r+85) \mu _{\pi }{}^2+2 \left(2 \bar{Q}^2 (8 r+23)-3 r (r (34 r+65)-93)-839\right)\right)\nonumber\\
&+&r^2\,{\cal C}_{\Delta\Delta}\,\int_0^1 dx\int_0^{1-x} dy\,\Big\{\Big[y (1-x-y) \left(6 x^4+2 \left(2 \left(r^2+r+\bar{Q}^2+1\right)-\left(5 \bar{Q}^2+(r\right.\right.\right.\nonumber\\
&-&\left.\left.\left.1) (5 r+7)\right) y\right) x^3+\left(2 \left(2 \bar{Q}^4+(r (4
   r+3)-11) \bar{Q}^2+(r-1)^2 (r+1) (2 r+5)\right) y^2\right.\right.\nonumber\\
&+&\left.\left.\left(-3 \bar{Q}^4
+(9-2 r (3 r+4)) \bar{Q}^2-(r-1) (r+3) (r (3 r+2)+4)\right) y+2 \left(\bar{Q}^2 (r+1)\right.\right.\right.\nonumber\\
&+&\left.\left.\left.r
   (r (r+2)+2)
-2\right)\right) x^2-\left((r (y-1)-4 y+3) y (2 y-1) \bar{Q}^4+\left(-9 y+2 \left(y (y (2 y-3)\right.\right.\right.\right.\nonumber\\
&+&\left.\left.\left.\left.1) r^3+y ((6-5 y) y
-2) r^2+\left(y
   \left(-4 y^2+2 y+3\right)-1\right) r+y^2 (7 y+1)\right)+2\right) \bar{Q}^2\right.\right.\nonumber\\
&+&\left.\left.(r-1) \left(2 \left(r^2-1\right)^2 y^3-(r
-1) (r+1) \left(3
   r^2+r+8\right) y^2+\left(r \left(r^3+10 r+7\right)-6\right) y\right.\right.\right.\nonumber\\
&-&\left.\left.\left.2 \left(r^2+r+1\right)\right)\right) x-(y-1)
y \left(((2 r (y-1)-4 y+3)
   y-1) \bar{Q}^4+(r (y-1)-y) \left(r\right.\right.\right.\nonumber\\
&+&\left.\left.\left.2 \left(r^2-1\right) y-9\right) \bar{Q}^2+(r-1)^2
(r+1)\right)\right)\Big]\frac{1}{2 \bar{\mathcal{M}}^2_{\Delta \Delta }}\nonumber-\Big[\left((56 y-4) x^3+\left(-2 \left(22 \bar{Q}^2+r (22 r\right.\right.\right.\nonumber\\
&+&\left.\left.\left.17)-62\right) y^2+\left(20 \bar{Q}^2+4 r (5 r
+8)-31\right) y-2 (r+1)\right) x^2+2
   \left(\left(11 \bar{Q}^2 (r-4)\right.\right.\right.\nonumber\\
&+&\left.\left.\left.(r-1) (r (11 r-17)-40)\right) y^3+\left(\bar{Q}^2 (47
-13 r)+r ((30-13 r) r+11)+2\right) y^2+\left(\bar{Q}^2 (3 r\right.\right.\right.\nonumber\\
&-&\left.\left.\left.14)+r (r (3
   r-10)+5)-9\right) y-r+1\right) x
+y \left(12 (r-1)^2 (r+1) y^3-(r-1) (r (24 r\right.\right.\nonumber\\
&+&\left.\left.5)+31) y^2+r (4 r (4 r-3)+59) y-32 y+r ((3
-4 r) r-21)+2 \bar{Q}^2
   (y ((37-22 y) y-22)\right.\right.\nonumber\\
&+&\left.\left. (y-1) (y (11 y-10)+2)+5)+1\right)\right)\Big]
\frac{\log \left(\bar{\mathcal{M}}^2_{\Delta \Delta }\right)}{2}
+(7 x+32 y-50 y (x+y)+r (y (25 y\nonumber\\
&-&21)+2)-4) \bar{\mathcal{M}}^2_{\Delta \Delta }\log \left(\bar{\mathcal{M}}^2_{\Delta \Delta }\right)\Big\}.\nonumber
\end{eqnarray}
\end{document}